\documentclass[11pt,a4paper]{article}
\pdfoutput=1
\usepackage{jheppub}
\usepackage{rotating}
\usepackage{array}
\usepackage{amsmath}
\usepackage[normalem]{ulem}
\usepackage{slashed}
\usepackage{booktabs}
\usepackage[pdftex,table]{xcolor}
\usepackage{units}
\usepackage{xfrac}
\usepackage{mathtools}
\usepackage{empheq}

\usepackage{multirow}

\preprint{DESY-17-211}

\title{BBN constraints on MeV-scale dark sectors. \\ Part I. Sterile decays}

\author{Marco Hufnagel,}
\author{Kai Schmidt-Hoberg}
\author{and Sebastian Wild}

\affiliation{DESY, Notkestra\ss e 85, D-22607 Hamburg, Germany}

\emailAdd{marco.hufnagel@desy.de}
\emailAdd{kai.schmidt-hoberg@desy.de}
\emailAdd{sebastian.wild@desy.de}

\abstract{We study constraints from Big Bang Nucleosynthesis on inert particles in a dark sector which contribute to the Hubble rate and therefore change the predictions of the primordial nuclear abundances. We pay special attention to the case of MeV-scale particles decaying into dark radiation, which are neither fully relativistic nor non-relativistic during all temperatures relevant to Big Bang Nucleosynthesis. As an application we discuss the implications of our general results for models of self-interacting dark matter with light mediators.}


\begin{document}
\maketitle

\section{Introduction}

Among the various probes of the Standard Models of particle physics and cosmology, Big Bang Nucleosynthesis (BBN) stands out by dating further back in time than any other existing observation. Using the baryon-to-photon ratio inferred from the Cosmic Microwave Background (CMB), $\eta_\text{CMB}$, the predicted abundances of nuclei such as deuterium and helium which have been formed in the first minutes after the Big Bang are in excellent agreement with observations, in the case of deuterium to a precision of about one percent~\mbox{\cite{2016RvMP...88a5004C,Olive:2016xmw}}.

The primordial nuclear abundances depend quite sensitively on the Hubble rate $H$ during BBN, mainly due to the effect of the expansion rate on the temperature at which protons and neutrons fall out of equilibrium. The Hubble rate is fully determined by the total energy density $\rho$ which can receive contributions from particles appearing in theories beyond the Standard Model (SM). BBN is therefore a potentially powerful probe of scenarios that predict a significant extra energy density at $T \simeq 0.01-10.0\,\text{MeV}$~\cite{Shvartsman:1969mm,Steigman:1977kc,Scherrer:1987rr,1988ApJ...331...33S}.

This is particularly interesting in extensions of the SM containing a \emph{dark sector}, i.e.~a collection of particles which are only weakly coupled to (or even fully decoupled from) the SM heat bath. Due to the small couplings of the new states to SM particles, such scenarios are often hard to probe at terrestrial experiments such as colliders or dark matter direct detection experiments. However, as long as the temperature of the dark sector is not much smaller than the photon temperature, even a fully decoupled dark sector contributes significantly to the energy budget of the Universe and hence alters the predictions of BBN. If the masses of the relevant particles in the dark sector are sufficiently below the MeV scale, $m_X \ll \text{MeV}$, they have essentially the same impact on the nuclear abundances as extra SM-like neutrinos. In this case, their effect can be fully described by a single number, $\Delta N_\text{eff}$, defined as the number of neutrinos leading to the same energy density as the corresponding dark sector particles. Using the most recent observations of primordial nuclear abundances as well as 
information from the Cosmic Microwave Background (CMB), the current upper limit is given by $\Delta N_\text{eff} \lesssim 0.2$ at 95\% C.L.~\cite{2016RvMP...88a5004C}. Furthermore, for the case of a stable or decaying dark sector particle $X$ which is fully \emph{non-relativistic} during BBN (i.e.~$m_X \gtrsim 10 - 100\,$MeV), upper bounds on the mass density of $X$ have been derived in~\cite{Scherrer:1987rr,1988ApJ...331...33S,Menestrina:2011mz}.

On the other hand, particles with $ 0.01 \,\text{MeV} \lesssim m_X \lesssim 10\,$MeV are neither ultra- nor non-relativistic for all temperatures relevant to BBN, and hence their effect on the predicted nuclear abundances cannot be simply parameterised by a temperature-independent number such as $\Delta N_\text{eff}$. Dark sector particles with masses in the MeV range actually appear in various extensions of the Standard Model, e.g.\ in the form of dark photons or dark Higgs bosons. Interestingly these particles can mediate the interactions of dark matter (DM), and if the DM particle is sufficiently heavier, an MeV-scale mediator can lead to significant velocity dependent DM self-interactions~\cite{Ackerman:mha,Feng:2009mn,Buckley:2009in,Feng:2009hw,Loeb:2010gj,Aarssen:2012fx,Tulin:2013teo,Kaplinghat:2015aga}, which in turn may solve potential small-scale structure problems of the collisionless cold dark matter paradigm (see \cite{Tulin:2017ara} for a recent review). 

In this work, we perform for the first time a dedicated study of BBN constraints on such a scenario, i.e.\ an MeV-scale dark sector particle which is either stable or decays into other dark sector states while being thermally decoupled from the SM heat bath. We outline our procedure for calculating the relevant nuclear abundances and how we derive the bounds from BBN in the next section. In section~\ref{sec:evolution} we show how the Hubble rate evolves in our general setup and evaluate the ensuing constraints from BBN. In section~\ref{sec:model} we apply our general results to a specific model, which features dark matter self-interactions as well as late kinetic decoupling and has therefore been proposed as a solution to the possible small scale structure problems mentioned above. In section~\ref{sec:conclusions} we summarise our results. Some technical details of our calculations can be found in appendices~\ref{sec:impact_kineticdecoupling} and \ref{sec:freezein}.

\section{Calculation of nuclear abundances and comparison to observations}
\label{sec:calc_nuclabundances}

For a given particle content in the dark sector, the relevant quantity affecting primordial nuclear abundances is the corresponding additional energy density $\rho_D(T)$ on top of the SM contribution $\rho_\text{SM}(T)$, leading to a modified Hubble rate
\begin{align}
H(T) = \left[ \frac{8 \pi G}{3} \left( \rho_\text{SM}(T) + \rho_D(T) \right) \right]^{1/2}  \,,
\label{eq:Hubble_rate}
\end{align}
where $G$ is Newton's constant. An increased Hubble rate during the era of BBN modifies the temperature at which neutrons and protons fall out of equilibrium and hence changes the amount of neutrons available for the formation of helium \cite{PhysRevLett.16.410,Shvartsman:1969mm}. In addition the abundances of heavier nuclei, which are formed at later times, depend on the interplay between the corresponding reaction rates and $H(T)$ and are hence also modified by the existence of the dark sector. We emphasise that for a fully decoupled dark sector as discussed in this work, the change in the Hubble rate is in fact the \emph{only} source of non-standard BBN physics, as by definition none of the particles in the dark sector directly interact with nuclei.

Before discussing the derivation of $\rho_D(T)$ for specific setups in sections~\ref{sec:evolution} and~\ref{sec:model}, let us now first describe our procedure of calculating primordial nuclear abundances for a given Hubble rate defined via eq.~(\ref{eq:Hubble_rate}). We compute the primordial abundances of $\text{H}$, ${}^2\text{H}$, ${}^3\text{He}$, ${}^6\text{Li}$ and ${}^7\text{Li}$ using a modified version of the public {\tt AlterBBN} code~\cite{Arbey:2011nf} (see also~\cite{Riemer-Sorensen:2017vxj}), in which we replace the standard expression for the Hubble rate by eq.~(\ref{eq:Hubble_rate}). The predicted abundances are then compared to the most recent compilation of observations~\cite{Olive:2016xmw}:\begin{align}
& Y_\text{p} \quad & (2.45 \pm 0.04) \times 10^{-1} \\
& \text{D}/\text{H} \quad & (2.53 \pm 0.04) \times 10^{-5} \\
& ({}^6\text{Li} + {}^7\text{Li})/\text{H} \quad & (1.6 \pm 0.3) \times 10^{-10} \\
& {}^6\text{Li}/{}^7\text{Li} \quad & \lesssim 0.05
\end{align}
As is well known there is a discrepancy between the SM prediction of the amount of lithium and the value inferred from astrophysical observations \cite{Fields:2011zzb}. While there is a slight alleviation of this discrepancy for a larger Hubble rate as in our setup, we have confirmed that when taking into account constraints from other elements, the change in the lithium abundance is small and does not significantly improve the result compared to the SM case. As the inference of the \emph{primordial} lithium abundance suffers from large astrophysical uncertainties, in particular due to stellar depletion~\cite{Korn:2006tv}, we conservatively only take into account the limits from helium and deuterium in the following.

It is well known that due to the ever more precise observations of primordial abundances (in particular of deuterium), theoretical errors associated to nuclear reaction rates have become an important or sometimes even dominant source of systematic uncertainty of BBN constraints on physics beyond the SM. 
The $\pm 1 \sigma$ high and low values of the nuclear reaction rates are implemented in {\tt AlterBBN} and we compute the corresponding abundance ratios $R_i$, $R_i^{+\sigma}$ and $R_i^{-\sigma}$.  
We then define the theoretical error on each abundance ratio via
\begin{equation}
\sigma_{R_i}^{\text{th}} = \min_i\left(\left|R_i - R_i^{+\sigma}\right|, \left|R_i- R_i^{-\sigma}\right|\right) \,,
\end{equation}
i.e.~in case of asymmetric errors we choose the smaller one as a proxy for a one sigma Gaussian error\footnote{This procedure gives similar results compared to the method developed in \cite{Fiorentini:1998fv}.}. A given point in parameter space is then considered to be excluded at the $n$-sigma level if
\begin{equation}
\Delta_{R_i} \equiv \left| R_i - R_i^{\text{obs}} \right| \bigg/ \sqrt{ \left( \sigma_{R_i}^{\text{th}} \right)^2 + \left( \sigma_{R_i}^{\text{obs}} \right)^2 } \geq n
\label{eq:Delta_Ri}
\end{equation}
for at least one abundance ratio $R_i$.

    For a given exotic energy density, the calculation of the ratios $R_i$ further depends on the neutron lifetime $\tau_n$ as well as the baryon-to-photon ratio $\eta$. In the following we will set the neutron lifetime to the PDG value of $\tau_n = 880\;\mathrm{s}$~\cite{2016RvMP...88a5004C} (we have checked that a variation of the neutron lifetime within its uncertainties does not appreciably affect our results). The baryon-to-photon ratio can either be left as a free parameter which is then constrained by BBN observations themselves, or by using additional information, e.g.\ by employing the measurements from the CMB. The first option may give rise to a sizeable range 
of possible values for $\eta$ during BBN, e.g.~we obtain $\eta_\text{BBN} \in [5.9, 6.4]\times 10^{-10}$ assuming SM physics (in sufficient agreement with~\cite{Olive:2016xmw}). Employing the most recent CMB measurement fixes $\eta_\text{CMB} = (6.10 \pm 0.04)  \times 10^{-10} $~\cite{Ade:2015xua} and therefore leads to more stringent BBN constraints. It should be kept in mind that, by deriving constraints from BBN using the determination of the baryon-to-photon ratio from the CMB, one implicitly assumes that $\eta$ is constant between $T_\text{BBN} \simeq \text{MeV}$ and $T_\text{CMB} \simeq \text{eV}$, that is $\eta_\text{BBN}=\eta_\text{CMB}$. While this is indeed the case for the scenario of a light mediator coupled to a sterile neutrino discussed later in this work, this assumption can be violated in other set-ups with entropy production after the end of BBN. In light of this, we derive two sets of exclusion limits denoted by \emph{BBN only} and \textit{BBN+$\eta_{\text{CMB}}$}. In the former case, we vary $\eta$
and exclude a data point only in case the criterion~(\ref{eq:Delta_Ri}) is fulfilled for all baryon-to-photon ratios. In the latter case, we explicitly fix $\eta$ to the central CMB value\footnote{The value of $\eta$ inferred from the CMB is largely independent of the extra energy density~\cite{2016RvMP...88a5004C}, so it is a good approximation to use the value $\eta_\text{CMB}$ inferred under the assumption of SM physics.} and exclude a data point if the criterion~(\ref{eq:Delta_Ri}) is fulfilled for this particular baryon-to-photon ratio. 

As a validation of our procedure to constrain additional energy densities using BBN observations, we re-derive the standard constraints on the effective number of additional neutrinos $\Delta N_\text{eff}$, corresponding to an additional energy density $\rho_D(T) \equiv \Delta N_\text{eff} \cdot 2 \cdot \frac78 \frac{\pi^2}{30} T_\nu(T)^4$, with $T_\nu(T)$ being the temperature of the SM neutrinos as a function of the photon temperature $T$. At $2\sigma$, $3\sigma$ and $5\sigma$, we obtain a \textit{BBN+$\eta_{\text{CMB}}$} upper limit of $\Delta N_\text{eff} \simeq 0.31, 0.54, 1.02$, respectively, which is in good agreement with other recent results~\cite{Olive:2016xmw}.

\section{Generic constraints for a decoupled MeV-scale dark sector particle}
\label{sec:evolution}

\subsection{Setup and assumptions}

Having defined our procedure of obtaining BBN bounds for a given $\rho_D(T)$, we now proceed to the actual calculation of the exotic energy density in a fairly general setup of a decoupled MeV-scale particle. More precisely, we consider an extension of the SM by a dark sector which in particular contains an additional boson $\phi$ with $m_{\phi}\sim\text{MeV}$ as well as a light particle $N$ with a mass well below the $\text{MeV}$ scale\footnote{At this point there is no dependence on the quantum numbers of the light state.}. For studying generic BBN constraints on such a scenario it is not necessary to specify the complete particle content of the dark sector\footnote{In fact, given that we work with a minimal particle content, the dark sector energy density will be minimal and our bounds will be correspondingly conservative.}; hence, we will only make the following minimal assumptions regarding its cosmological history:
\begin{itemize}
\item At large temperatures $T$, the particle $\phi$ is in chemical equilibrium within the dark sector and then decouples at a photon temperature $T_{\text{cd}}$, corresponding to the temperature $T_{D}(T_\text{cd})$ in the dark sector. This setup could e.g.~be realised by coupling $\phi$ to an additional dark matter particle $\psi$ which freezes out via the process $\psi \bar \psi \leftrightarrow \phi \phi$.
\item $\phi$ can decay into $N \bar N$, with the lifetime $\tau_\phi$ being a free parameter to be constrained by BBN observations.
\item The visible and hidden sector are fully decoupled, i.e.~any couplings of $\phi$ and $N$ to SM states are considered to be sufficiently small.
\item If not stated otherwise, we also assume that $N$ never thermalises within the dark sector, implying in particular a small coupling to $\phi$, corresponding to large lifetimes $\tau_\phi$. However, we will explicitly abandon this assumption in section~\ref{sec:model} where we discuss the relevance of BBN constraints in the context of a simple model of dark matter coupled to $\phi$.
\end{itemize}
With these assumptions, there are four free parameters relevant for the discussion of BBN constraints: the temperature $T_{\text{cd}}$ of the SM sector at which the particle $\phi$ decouples, the corresponding temperature $T_\text{D}(T_\text{cd})$ of the dark sector, the mass $m_\phi$ and the lifetime $\tau_\phi$ of the MeV-scale particle $\phi$. For given values of these parameters, we now outline the derivation of the energy densities $\rho_\phi(T)$ and $\rho_N(T)$ at $T < T_\text{cd}$, which enter the Hubble rate $H(T)$ defined in eq.~(\ref{eq:Hubble_rate}) and hence the calculation of the primordial abundances via\footnote{This is largely independent of the energy density in dark matter, which is typically negligible at the time of BBN.} $\rho_D(T) \simeq \rho_\phi(T) + \rho_N(T)$.

\subsection{Cosmological evolution of the dark sector}

After chemical and kinetic\footnote{We note that after chemical decoupling $\phi$ could still be in kinetic equilibrium with the dark sector down to $T_\text{kd} < T_\text{cd}$, leading to a collision operator $\mathcal{C}_\text{elastic}[f_\phi]$ appearing on the right hand side of eq.~(\ref{eq:Boltzmann_eqn_phi}). However, we show in appendix~\ref{sec:impact_kineticdecoupling} that the impact of $T_\text{kd}$ on our results is negligible and consequently we set $T_\text{kd} = T_\text{cd}$ in the following discussion. } decoupling of $\phi$ from the dark sector heat bath at $T = T_\text{cd}$, the Boltzmann equation for the phase space density $f_\phi(t,E)$~\cite{Kolb:1990vq} rewritten in terms of the photon temperature $T$ and the momentum of $\phi$ $p$ is given by
\begin{align}
\left( \frac{T}{1+\Delta_{*s}(T)} \frac{\partial}{\partial T} + p \frac{\partial}{\partial p} \right) f_\phi(T,p) &= \frac{1}{\tau_\phi} \frac{m_\phi}{H(T)\sqrt{m_\phi^2 + p^2}}f_\phi(T, p)\;, \label{eq:Boltzmann_eqn_phi} 
\end{align}
with $\Delta_{*s}(T) \equiv T/(3 g_{*s}(T)) \cdot \text{d} g_{*s}(T)/\text{d} T$ and where $g_{*s}(T)$ denotes the number of entropy degrees of freedom of the SM.
Furthermore, at chemical decoupling we have 
\begin{align}
f_\phi(T = T_\text{cd},p) &= \bar{f}_\phi(T = T_\text{cd},p)\;\,, \nonumber
\end{align}
with $\bar{f}_\phi(T,p)$ being the phase space distribution of $\phi$ in chemical equilibrium. The solution of this initial value problem is given by
\begin{empheq}{align}
f_\phi (T,p)\big|_{T < T_\text{cd}} = & \left[\;\, \exp{\left( \frac{\sqrt{m_\phi^2 + p_{*}^2(T, T_{\text{cd}})}}{T_D(T_{\text{cd}})} \right)} - 1 \;\,\right]^{-1} \nonumber \\
& \times \exp{\left( - \frac{1}{\tau_\phi} \int_{T}^{T_{\text{cd}}} \frac{m_\phi}{\sqrt{m_\phi^2 + p_{*}^2(T, \lambda)}} \cdot \frac{1+\Delta_{*s}(\lambda)}{H(\lambda)\lambda} \; \text{d}\lambda \right)}
\label{phi_after_cd}
\end{empheq}
with the red-shifted momentum
\begin{equation}
p_{*}^2(T, \lambda) \equiv p^2\cdot \left( \frac{g_{*s}(\lambda)^{\sfrac13}\lambda}{g_{*s}(T)^{\sfrac13}T} \right)^2\;\,.
\label{eq:p2star}
\end{equation}
This form of $f_\phi (T, p)$ is straightforward to understand: For $\tau_\phi \rightarrow \infty$, the exponential term vanishes and the phase-space density simply follows from redshifting all momenta according to $p \propto 1/R \propto g_{*s}(T)^{\sfrac13}T$, as defined in eq.~(\ref{eq:p2star}). For finite $\tau_\phi$, the abundance of the mediator is exponentially suppressed by a factor $\sim \exp(-t/(\gamma_\text{eff} \tau_\phi))$, where $\gamma_\text{eff}$ is the time-averaged (or equivalently temperature-averaged) Lorentz boost of the mediators. The energy and number density of $\phi$ then follow from
\begin{align}
\rho_\phi (T) &= \frac{g_\phi}{2 \pi^2} \int_{0}^{\infty} \text{d}p \, p^2 \sqrt{p^2 + m_\phi^2} \, f_\phi (T, p) \;\,,\label{rho_phi_f}\\
n_\phi (T) &= \frac{g_\phi}{2 \pi^2} \int_{0}^{\infty} \text{d}p \, p^2 \, f_\phi (T, p) \;\,,
\label{n_phi_f}
\end{align}
with $g_\phi$ denoting the number of spin-degrees of freedom of $\phi$. In the following we will discuss scalar ($g_\phi=1$) as well as massive vector ($g_\phi=3$) fields.

The decrease in $\rho_\phi$ due to the decay of the mediators $\phi$ is balanced by the increase of the energy density of the decay products $N$, which then also contributes to the expansion rate of the Universe. If the only production mechanism of $N$ is the decay of the particle $\phi$, the integrated Boltzmann equation for $\rho_N$ (i.e.~the summed energy density of $N$ and $\bar{N}$) takes the form
\begin{equation}
	\frac{\text{d}\rho_N(T)}{\text{d}T} - 4\frac{1+\Delta_{*s}(T)}{T} \rho_N(T) = - \frac{1+\Delta_{*s}(T)}{H(T)T} \frac{m_\phi n_\phi(T)}{\tau_\phi}\;, \quad \rho_N(T = T_\text{cd}) \simeq 0\;\,.
\label{eq:rhoN_eqn}
\end{equation}
Here, the second term on the left-hand side accounts for the decrease of $\rho_N$ due to the expansion of the universe ($\rho_N \propto (g_{*s}^{\sfrac13}T)^4$ without $\phi$ decay), while the right-hand side describes the increase of $\rho_N$ due to the decay of $\phi$, properly taking into account the Lorentz factor in the decay process. For given $n_\phi (T)$, the solution to eq.~(\ref{eq:rhoN_eqn}) is given by
\begin{align}
\rho_N(T) = \frac{1}{\tau_\phi} \left(g_{*s}(T)^{\sfrac13}T\right)^4  \int_{T}^{T_\text{cd}}  \frac{m_\phi n_\phi(\lambda)}{\left(g_{*s}(\lambda)^{\sfrac13}\lambda\right)^4} \cdot \frac{1+\Delta_{*s}(\lambda)}{H(\lambda)\lambda}\; \text{d}\lambda\;\,.
\label{rho_N+Nb}
\end{align}

We remark that in accord with eq.~(\ref{eq:Hubble_rate}), $H(T)$ also contains a contribution from dark sector particles, necessitating an iterative solution of eqs.~(\ref{phi_after_cd}) and~(\ref{rho_N+Nb}). We have confirmed however that for deviations of $H(T)$ relevant to BBN constraints, such a recursive procedure does not significantly change our results and that it is sufficient to only take into account the SM contribution to the Hubble parameter in eqs~(\ref{phi_after_cd}) and~(\ref{rho_N+Nb}).

\begin{figure}
\begin{center}
\hspace*{-0.3cm}
\includegraphics[scale=0.87]{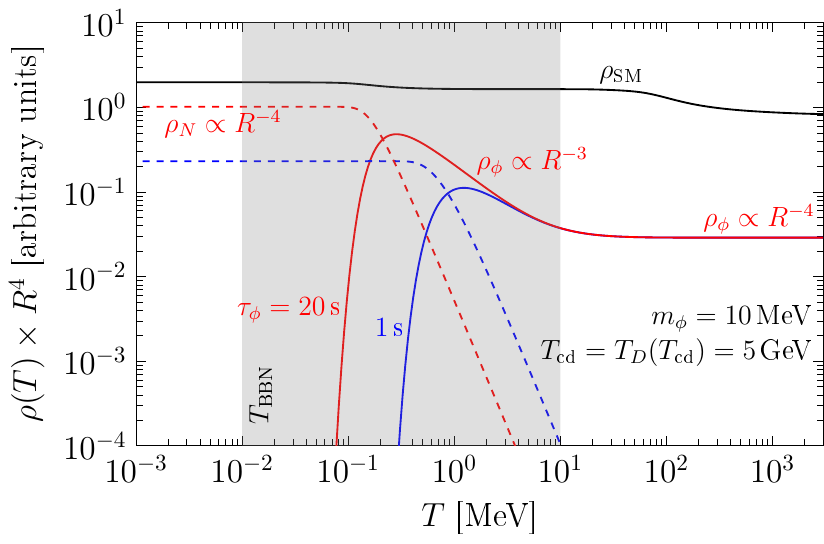}
\hspace*{0.2cm}
\includegraphics[scale=0.87]{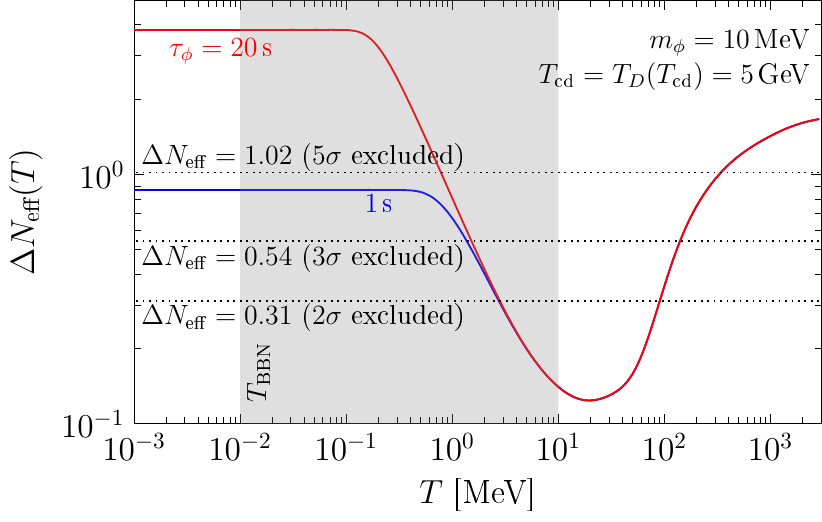}
\end{center}
\caption{\small \textit{Left panel:} evolution of the energy densities $\rho_\phi$ and $\rho_N$ for exemplary choices of $m_\phi$, $T_\text{cd}$, $T_D(T_\text{cd})$ and two different lifetimes of the particle $\phi$ as a function of the photon temperature $T$. For comparison we also show the energy density of the SM sector $\rho_\text{SM}$. \textit{Right panel:} temperature dependence of the equivalent effective number of additional neutrinos $\Delta N_\text{eff}(T)$ for the same scenarios. The temperatures relevant to BBN are indicated in grey. }
\label{fig:rhoT_Nefft_modelindependent}
\end{figure}

We illustrate the solutions for $\rho_\phi(T)$ and $\rho_N(T)$ for different choices of parameters in the left panel of Fig.~\ref{fig:rhoT_Nefft_modelindependent}. As long as $t \ll \tau_\phi$ and $T \gg m_\phi$ the energy density of $\phi$ simply follows the usual scaling behaviour for massless particles, $\rho_\phi \propto R^{-4} \propto g_{*s}(T)^{4/3} T^4$. For $T \lesssim m_\phi$, the energy density instead falls off less steeply ($\rho_\phi \propto R^{-3}$), leading to a relative increase of $\rho_\phi$ compared to the massless case. Furthermore, at temperatures corresponding to $t \gtrsim \tau_\phi$, the abundance of $\phi$ becomes exponentially suppressed and its energy density is transferred to the massless decay products $N$, which again scale as $\rho_N \propto R^{-4}$. Hence, the total energy density $\rho_D(T)$ increases with larger $m_\phi$ and/or $\tau_\phi$ and we can already anticipate that the BBN bounds on such a scenario will be strongest for sufficiently heavy and long-lived mediators.

In the right panel of Fig.~\ref{fig:rhoT_Nefft_modelindependent}, we show for the same set of model parameters the corresponding effective number of SM neutrinos which would give rise to the same energy density,
\begin{align}
\Delta N_\text{eff} (T) \equiv \frac{\rho_\phi(T) + \rho_N(T)}{2 \cdot \frac78 \frac{\pi^2}{30} T_\nu(T)^4} \,.
\end{align}
It can be seen from Fig.~\ref{fig:rhoT_Nefft_modelindependent} that for a decaying MeV-scale mediator $\Delta N_\text{eff} (T)$ can vary by more than one order of magnitude during the temperatures relevant to BBN, and the level of tension between predicted and observed nuclear abundances cannot be directly inferred by comparing to the corresponding exclusion limits on a constant value of $\Delta N_\text{eff}$.

\subsection{Resulting bounds from BBN}
\label{sec:generalBBN_bounds}

\begin{figure}
\begin{center}
\hspace*{-1.0cm}
\includegraphics[scale=0.43]{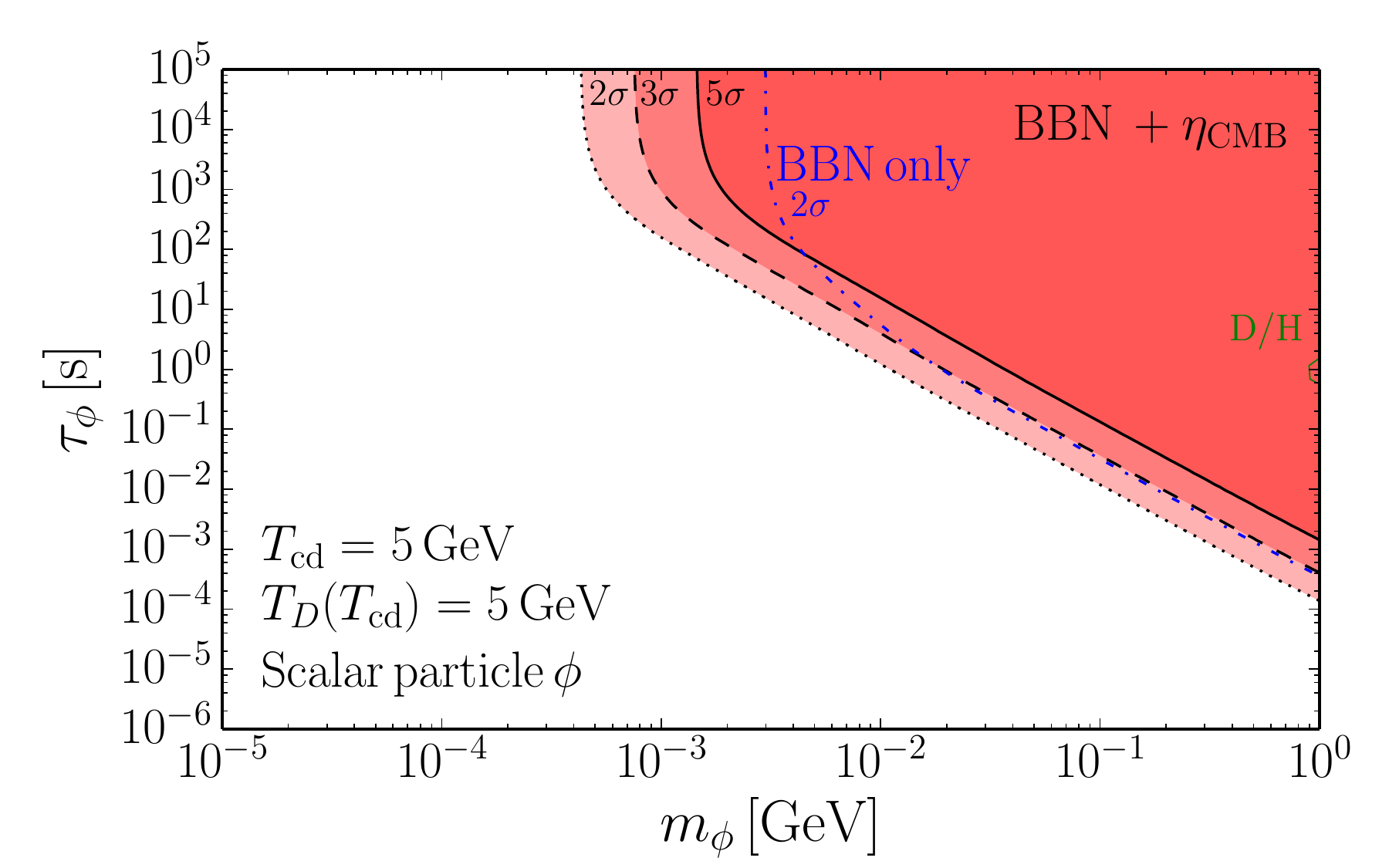}
\includegraphics[scale=0.43]{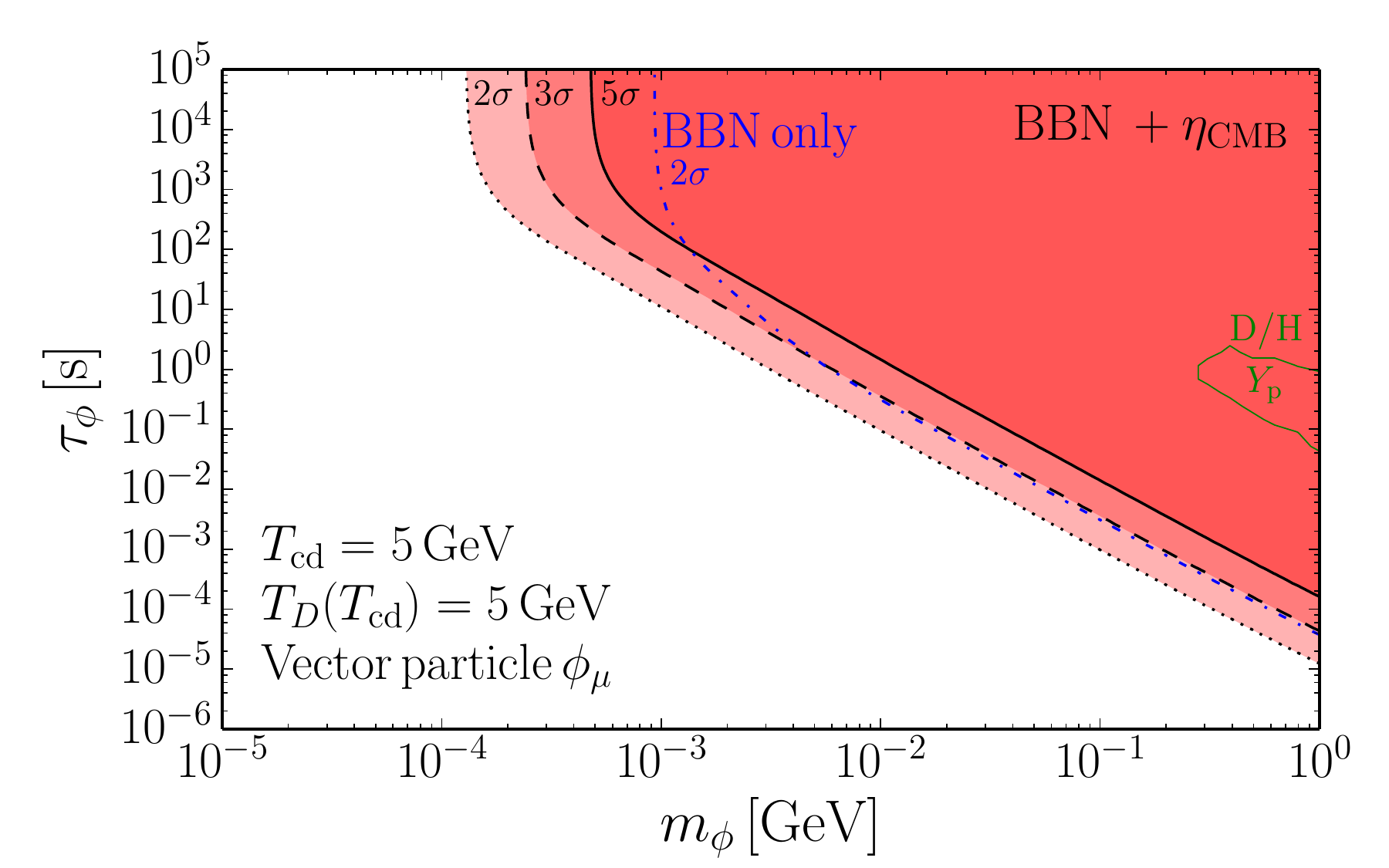}
\\
\hspace*{-1.1cm}
\includegraphics[scale=0.43]{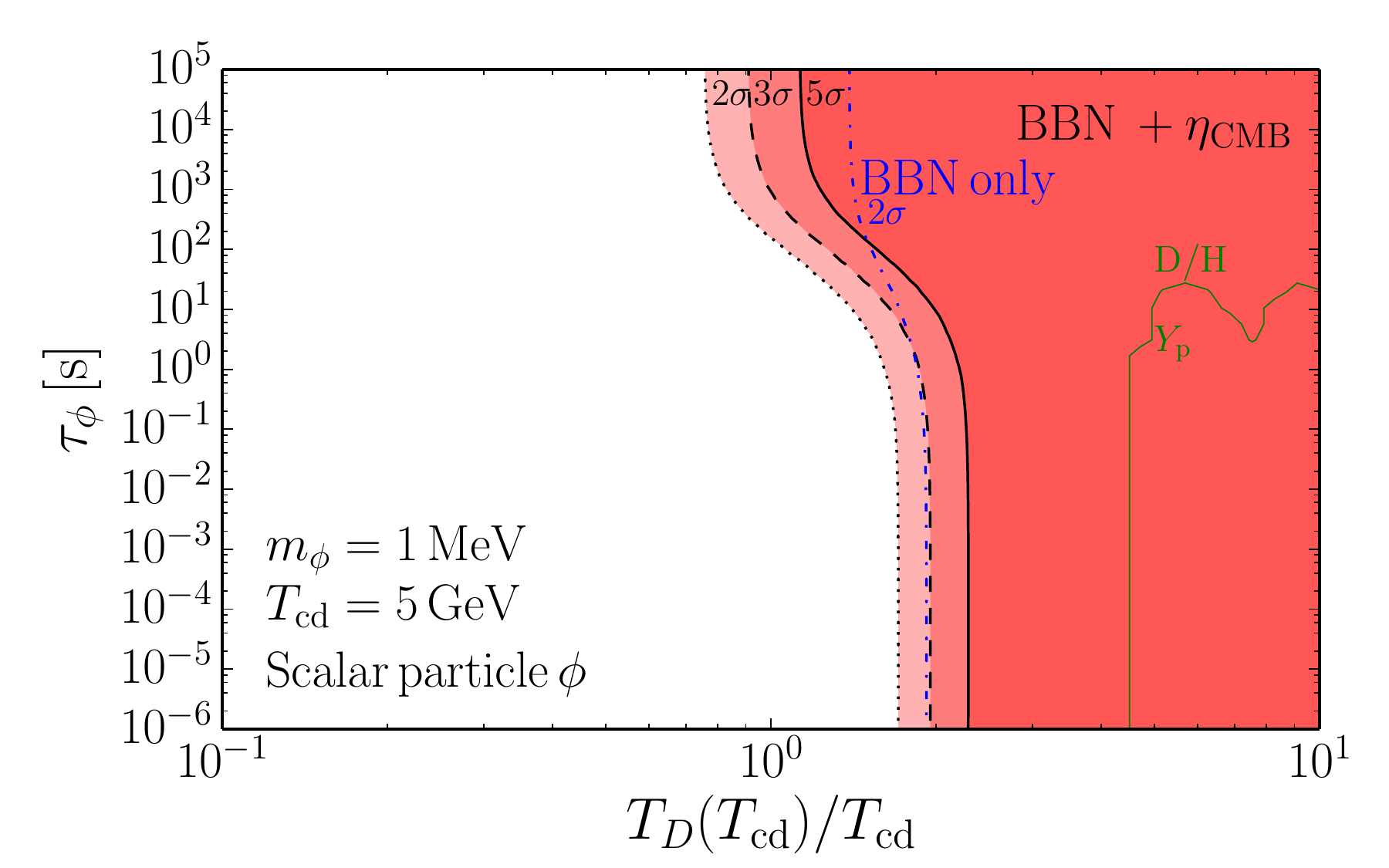}
\includegraphics[scale=0.43]{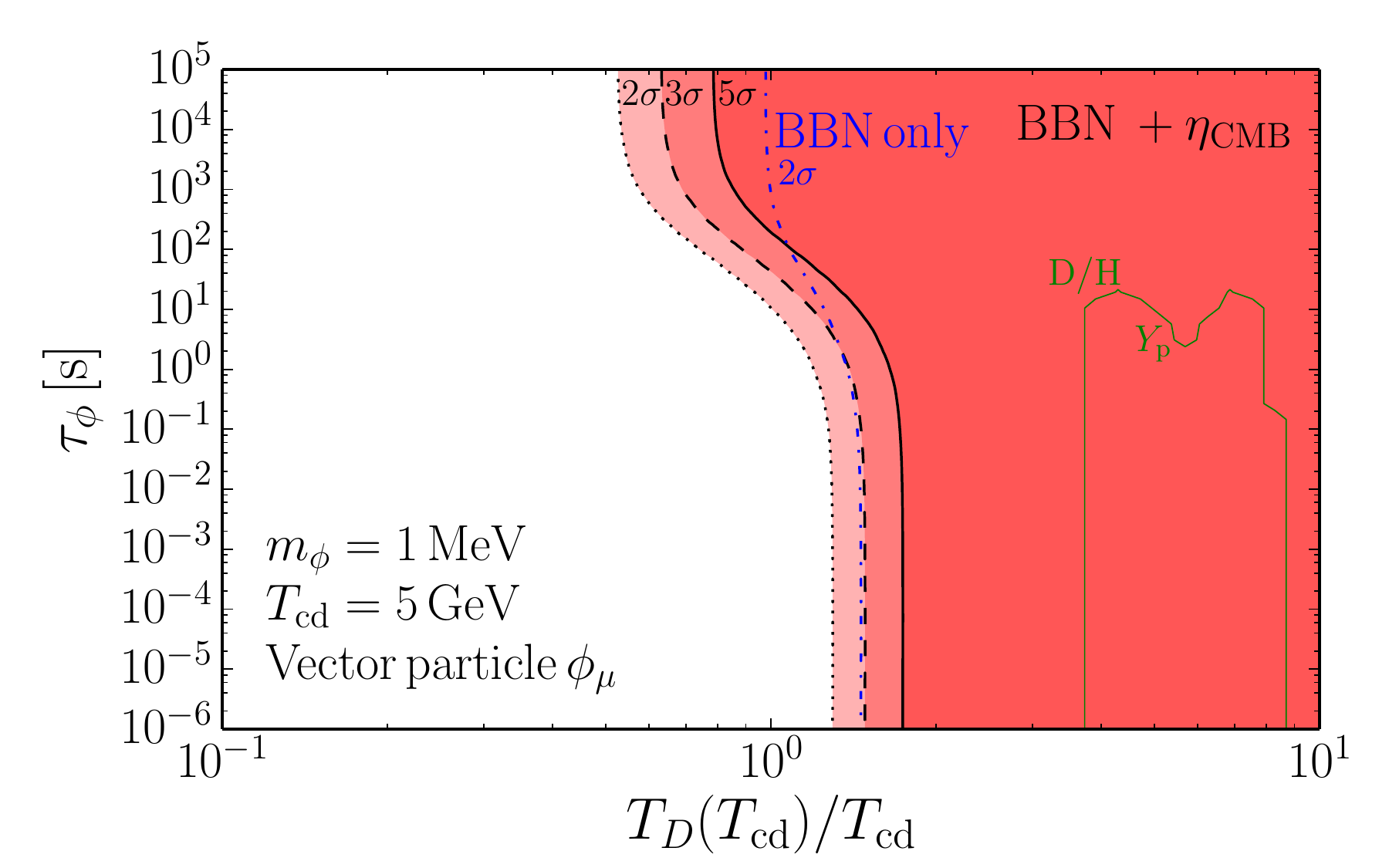}
\\
\hspace*{-1.1cm}
\includegraphics[scale=0.43]{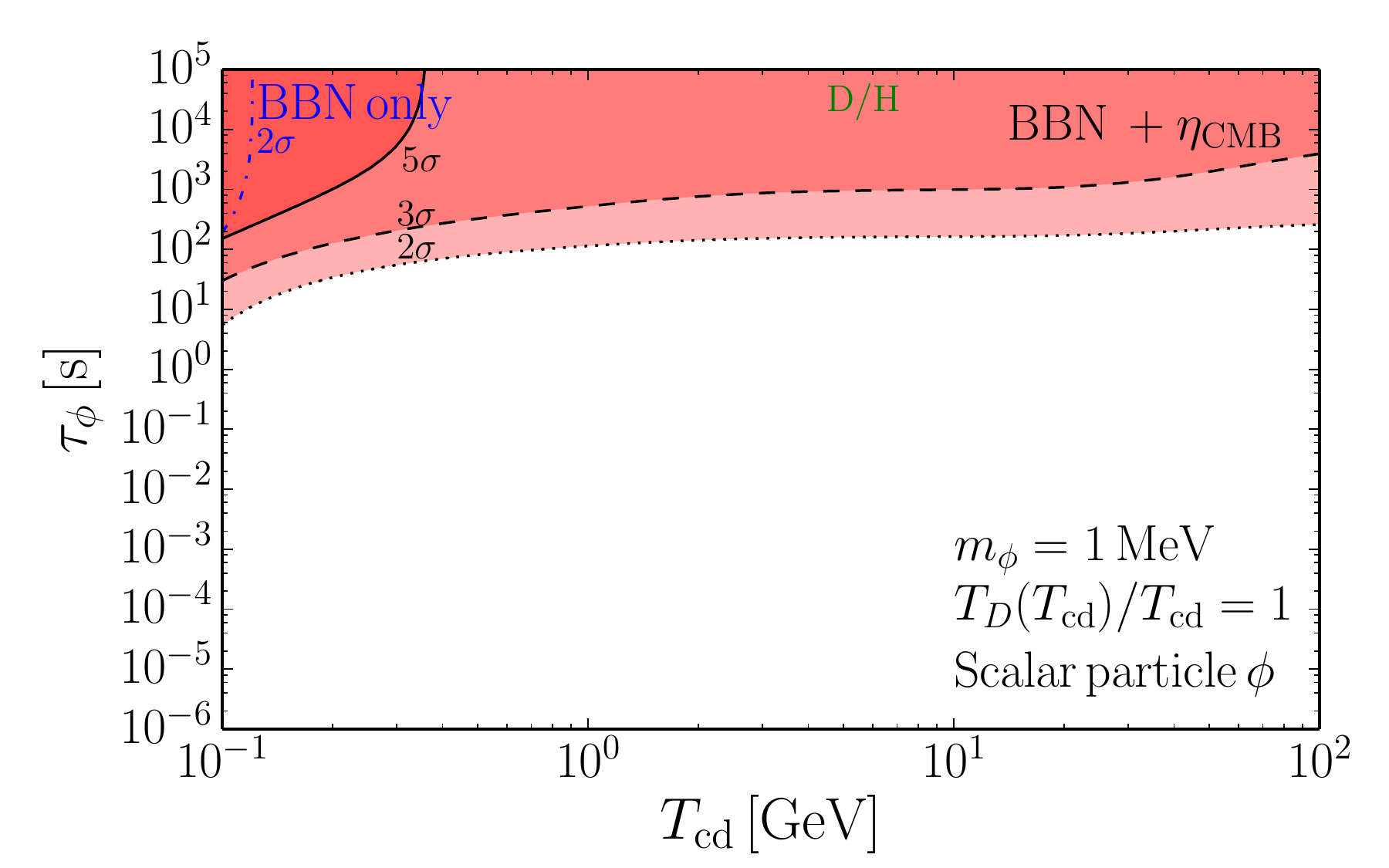}
\includegraphics[scale=0.43]{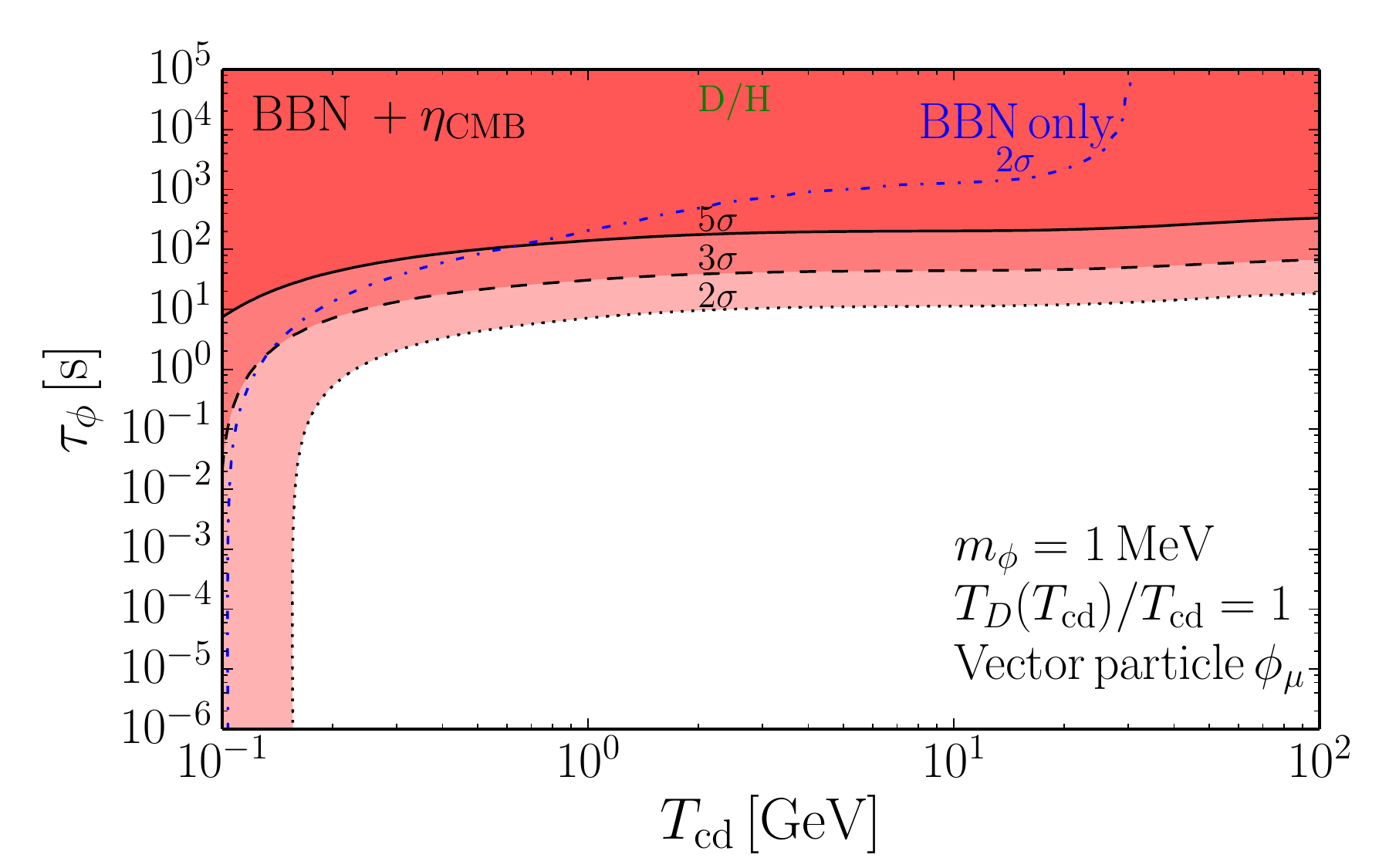}
\end{center}
\caption{\small BBN bounds for decaying scalar (\textit{left panels}) and vector (\textit{right panels}) particles in different slices of the parameter space. The red regions enclosed by the solid, dashed and dotted curves correspond to $5\sigma$, $3\sigma$ and $2\sigma$ bounds under the assumption $\eta_{\rm BBN}=\eta_{\rm CMB}$. 
The dash-dotted blue curve denotes the $2\sigma$ \emph{BBN only} bound, while the solid green curve indicates which nuclear abundance is mainly responsible for the exclusion.}
\label{fig:general_results}
\end{figure}

In this section we present the bounds from BBN on our scenario. As discussed above there are four relevant parameters, the mediator mass $m_\phi$ and lifetime $\tau_\phi$ as well as the chemical decoupling temperature $T_\text{cd}$ and the corresponding temperature in the dark sector $T_D(T_\text{cd})$. 
In Fig.~\ref{fig:general_results} we show the resulting bounds from BBN for scalars (left panels) and vectors (right panels), for different slices of the parameter space. The red regions indicate the $2\sigma, 3\sigma$ and $5 \sigma$ exclusions, making use of $\eta_\text{BBN}=\eta_\text{CMB}$. For comparison we also show a $2 \sigma$ bound for freely varying $\eta$, denoted \emph{BBN only}, which as expected is somewhat weaker. We further indicate whether the leading bound is due to the abundance of deuterium ($\text{D}/\text{H}$) or helium ($\text{Y}_\text{p}$).

The top panels show the dependence on the particle mass $m_\phi$ and lifetime $\tau_\phi$.
For sufficiently small $\tau_\phi$ and $m_\phi$, $\phi$ decays before BBN and while it is still relativistic. In that case, the energy density behaves as extra radiation throughout, corresponding to a constant $\Delta N_\text{eff}$ during BBN. For $T_\text{cd} = T_D(T_\text{cd}) = 5\,$GeV, the corresponding value can be easily obtained from
\begin{align}
\Delta N_\text{eff} \simeq \frac{g_\phi}{2 \cdot 7/8} \, \left( \frac{g_{\star S}(10\,\text{MeV})}{g_{\star S}(5\,\text{GeV})} \right)^{4/3} \simeq \begin{cases} 0.036 &\mbox{for a scalar }\phi  \\
0.11 & \mbox{for a vector }\phi \end{cases}
\end{align}
which as expected agrees with our numerical result of $\rho_D(T)$ in this region of parameter space. Clearly, this value of $\Delta N_\text{eff}$ is not excluded by BBN. Note that the resulting $\Delta N_\text{eff}$ is much smaller than unity largely because of significant entropy production in the SM sector between freeze out and BBN, increasing the photon temperature $T$ with respect to the dark sector temperature $T_D(T)$. Towards larger masses and lifetimes the dark sector energy density increases. The diagonal exclusion lines are essentially determined by the condition $T(t = \tau_\phi) \sim m_\phi$: for values of $\tau_\phi$ and/or $m_\phi$ larger than that, $\phi$ decays only after it has become non-relativistic. It then profits from the $R^{-3}$ scaling of non-relativistic matter as opposed to the $R^{-4}$ scaling of radiation, increasing $\rho_D(T)$ towards larger masses and lifetimes, leading to conflicts with observations. We furthermore observe that the bounds from BBN become insensitive to the lifetime of $\phi$ once $\tau_\phi$ is sufficiently large, as in this limit the particle is practically stable during BBN.

The panels in the second row show the dependence of the BBN bounds on the dark sector temperature $T_D(T_\text{cd})$ and the lifetime $\tau_\phi$. Increasing $T_D(T_\text{cd})$ compared to the photon temperature $T$ obviously increases the energy density in the dark sector, $\rho_D(T)$, and therefore leads to more stringent limits. Again the bounds get stronger for larger lifetimes when the particle $\phi$ becomes non-relativistic prior to its decay.

The bottom panels show the dependence on the chemical decoupling temperature $T_\text{cd}$ and lifetime $\tau_\phi$. It can be seen that the BBN bounds are largely insensitive to $T_\text{cd}$, except for small temperatures close to the QCD phase transition, where $g_{\star S}$ drops rapidly and the dark sector does not cool effectively compared to the SM sector any longer.

For $m_\phi \gtrsim 10\,$MeV, the decaying particle is non-relativistic for all temperatures relevant to BBN, a scenario discussed already in~\cite{1988ApJ...331...33S}. We conclude this section by updating the exclusion limits given in that work; the main improvement is that we take into account the most recent information on the baryon-to-photon ratio from the CMB as well as more recent input for the nuclear cross sections relevant to BBN. Following~\cite{1988ApJ...331...33S}, we parameterise the energy density of $\phi$ during BBN by $r = n_\phi(T_0)/n_\gamma(T_0)$, with $T_0 = 10^{12} \,\text{K} = 86.2\,$MeV being an arbitrary reference temperature prior to BBN. In the non-relativistic limit, the energy density of $\phi$ is simply given by $\rho_\phi(T) = m_\phi n_\phi(T)$, which in turn can be obtained from integrating the Boltzmann equation~(\ref{eq:Boltzmann_eqn_phi}):
\begin{equation}
\frac{\text{d}\rho_\phi(T)}{\text{d}T} - 3\frac{1+\Delta_{*s}(T)}{T} \rho_\phi(T) = \frac{1+\Delta_{*s}(T)}{H(T)T} \frac{\rho_\phi(T)}{\tau_\phi}\;, \quad \rho_\phi(T_0) = rm_\phi \cdot n_\gamma(T_0)\;\,.
\label{eq:rhophi_nonrel}
\end{equation}
Using $n_\gamma(T) = 2\zeta(3)T^3/\pi^2$, the solution to eq.~\eqref{eq:rhophi_nonrel} takes the form
\begin{equation}
\rho_\phi(T) = rm_\phi \cdot \frac{2\zeta(3)}{\pi^2} \frac{g_{*s}(T)T^3}{g_{*s}(T_0)} \exp\left(- \frac{1}{\tau_\phi} \int_{T}^{T_0} \frac{1+\Delta_{*s}(\lambda)}{H(\lambda)\lambda} \text{d}\lambda \right)\;\,.
\end{equation}
The energy density $\rho_N(T)$ of the decay products $N$ follows from eq.~(\ref{rho_N+Nb}) by setting\linebreak$T_{\text{cd}} \rightarrow \infty$ and $m_\phi n_\phi(\lambda) = \rho_\phi(\lambda)$.

\begin{figure}
	\centering
	\includegraphics[width=0.55\textwidth]{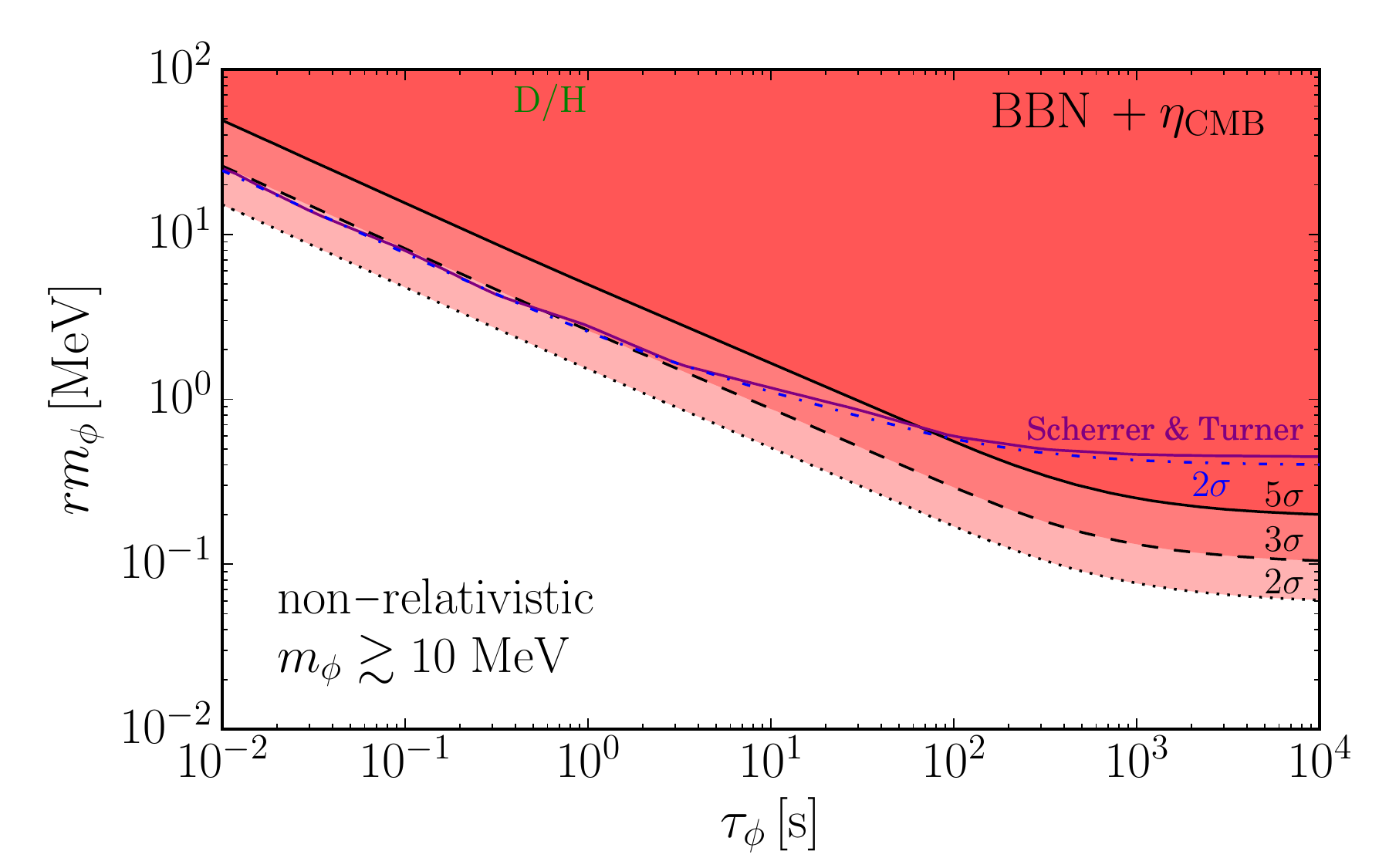}
	\caption{\small BBN bounds on non-relativistic decaying particles, valid for $m_\phi \gtrsim 10\,$MeV. Here, $r m_\phi$ parameterises the energy density of $\phi$ prior to its decay (see text for details). The black and blue curves are analogous to Fig.~\ref{fig:general_results}, while the solid purple curve shows the bound derived in~\cite{1988ApJ...331...33S}  for comparison.  }
	\label{fig:Scherrer_Turner}
\end{figure}

We show the resulting upper limits on $r m_\phi$ as a function of $\tau_\phi$ in Fig.~\ref{fig:Scherrer_Turner}. It can be seen that our \textit{BBN+$\eta_{\text{CMB}}$} bound improves significantly over the one presented in~\cite{1988ApJ...331...33S}, up to $\simeq 1$ order of magnitude in the limit of large lifetimes. Given that $m_\phi \gtrsim 10\,$MeV for this analysis to be valid, the plot in particular indicates that the number density of the extra particles has to be suppressed compared to the one of photons, $r \ll 1$, for lifetimes $\tau_\phi \gtrsim 1$~s. It can also be seen that our \emph{BBN only} bound closely follows the exclusion limit of~\cite{1988ApJ...331...33S}. In fact, our procedure of taking $\eta_\text{BBN}$ as a free parameter is similar to the approach in~\cite{1988ApJ...331...33S} undertaken prior to the precise CMB measurements of the baryon-to-photon ratio. However, let us remark that the overall agreement between these results is partially accidental: while we profit from much more precise measurements on the nuclear abundances, in contrast to~\cite{1988ApJ...331...33S} we take into account theoretical uncertainties in the nuclear reaction rates as discussed in section~\ref{sec:calc_nuclabundances}, leading again to weaker (but also more realistic) constraints.

\section{Implications for a model with dark matter self-interactions and late kinetic decoupling}
\label{sec:model}

\subsection{A model of dark matter coupled to a light mediator}

So far, the discussion of BBN bounds on MeV-scale decaying particles was largely model-independent, relying only on the assumptions specified at the beginning of section~\ref{sec:evolution}. In this section, we now explore the implications of these general constraints on more specific scenarios. We will be particularly interested in the case that the MeV-scale particle $\phi$ couples to dark matter $\psi$, naturally leading to sizeable dark matter self-interactions. The possibility of self-interacting dark matter has attracted an increasing amount of attention lately, as it may solve some tensions within the collisionless cold dark matter paradigm at small scales \cite{Spergel:1999mh}. Due to rather strong constraints on the dark matter self-scattering cross section in high velocity systems such as galaxy clusters \cite{Markevitch:2003at,Randall:2007ph,Peter:2012jh,Rocha:2012jg,Kahlhoefer:2013dca,Harvey:2015hha,Kaplinghat:2015aga}, a cross section which increases towards smaller velocities is preferred. Such a velocity dependence is most easily achieved with a light mediating particle \cite{Ackerman:mha,Feng:2009mn,Buckley:2009in,Feng:2009hw,Loeb:2010gj,Aarssen:2012fx,Tulin:2013teo,Kaplinghat:2015aga}. In such a setup the dark matter can naturally acquire its relic density via thermal freeze out with these mediators, $\psi \bar{\psi} \leftrightarrow \phi \phi$. These mediators on the other hand typically need to decay in order not to dominate the energy density of the Universe. It has been noted however that if the DM annihilation into the mediator is s-wave and the mediator decays into SM states, there are very strong reionisation bounds from the CMB and the parameter space leading to interesting dark matter self-scattering cross sections is excluded \cite{Bringmann:2016din,Cirelli:2016rnw}. A possibility to circumvent this conclusion would be to have the mediator decay into light hidden sector states such as sterile neutrinos, which do not lead to reionisation.\footnote{Other possibilities to circumvent this conclusion include a stable mediator which annihilates efficiently~\cite{Ma:2017ucp}, freeze-in production~\cite{Bernal:2015ova}, or more generally a sufficiently small dark sector temperature~\cite{Bringmann:2016din}. Scalar mediators lead to p-wave suppressed annihilation and are therefore unconstrained by CMB observations, but may suffer from a tension between direct detection experiments and constraints from BBN~\cite{Kaplinghat:2013yxa,Kainulainen:2015sva}, although a comprehensive analysis of BBN constraints remains to be done~\cite{BBNfuture}. Suppressing the scattering cross section relevant for direct detection allows to have viable models also for scalar mediators \cite{Blennow:2016gde,Kahlhoefer:2017umn}.} Interestingly such a model has been proposed as a solution to all small scale problems of the $\Lambda$CDM cosmology \cite{Aarssen:2012fx}, because in addition to DM self-interactions this model also allows for late kinetic decoupling, which suppresses structure at small scales and can therefore address the missing satellite problem\footnote{See however~\cite{Kim:2017iwr} for a recent critical re-evaluation of the missing satellite problem.}.

Following \cite{Aarssen:2012fx} let us discuss an example model featuring a Dirac dark matter particle $\psi$ coupled to a vector mediator $\phi_\mu$ which
in turn couples to light sterile neutrinos $N$ while couplings to the SM are suppressed. We also assume that mixing with active neutrinos is negligible. The relevant interaction terms are then given by
\begin{align}
\mathcal{L} = g_\psi \bar \psi \gamma^\mu \psi \phi_\mu + g_N \bar N \gamma^\mu N \phi_\mu \,.
\label{eq:lagrangian}
\end{align}
Assuming $m_N \ll m_\phi$, the vector mediator will then decay into a pair of sterile neutrinos with decay width
\begin{align}
\Gamma_\phi = \frac{g_N^2}{12 \pi} m_\phi \quad , \, \text{implying a lifetime} \quad \tau_\phi \simeq 2.5 \, \text{s} \, \cdot \left( \frac{g_N}{10^{-10}} \right)^{-2} \, \cdot \, \left( \frac{m_\phi}{\text{MeV}} \right)^{-1} \,.
\label{eq:Gammaphi}
\end{align}
Throughout the rest of this work, we will fix the coupling $g_\psi$ of dark matter to the mediator by the requirement of reproducing the observed relic density $(\Omega h^2)_\text{DM} \simeq 0.12$ via the freeze-out processes $\psi \bar{\psi}\leftrightarrow \phi \phi$ and (for sizeable values of $g_N$) $ \psi  \bar{\psi} \leftrightarrow N \bar N$. Besides $m_\psi$, $m_\phi$ and $g_N$, the value of $g_\psi$ leading to the correct relic density depends on the initial temperature ratio of the dark sector and the SM heat bath, which we denote as $(T_D/T)_\infty$. It is plausible to assume that at high energy scales additional interactions brought both sectors into thermal contact, corresponding to the choice $(T_D/T)_\infty = 1$. While this indeed will be our benchmark assumption, we will also discuss to what extent our results change when allowing for different values of the initial temperature ratio.
To proceed, we first determine the dark sector temperature $T_D(T)$ as a function of the photon temperature $T$ by demanding separate entropy conservation in the dark and visible sector, and then employ the semi-analytical approach for hidden sector freeze-out presented in~\cite{Feng:2008mu}\footnote{Notice that in eq.~(34) of~\cite{Feng:2008mu}, $\xi^{3/2}$ should be replaced by $\xi^{5/2}$.}.

\begin{figure}
	\centering
	\hspace*{-0.8cm}
	\includegraphics[scale=0.9]{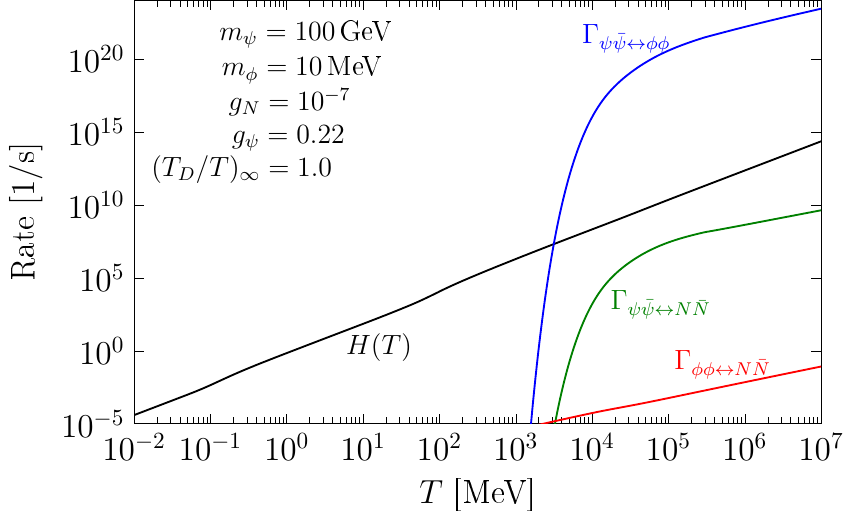}
	\includegraphics[scale=0.9]{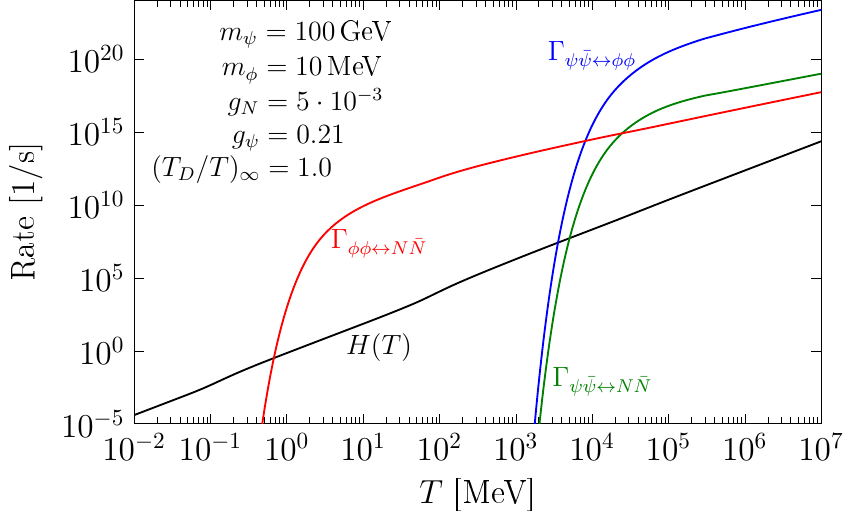}
	\caption{\small Comparison of the various annihilation rates $\Gamma$ to the Hubble rate $H$ for a value of $g_N$ corresponding to the small (\textit{left panel}) and large coupling regime (\textit{right panel}). In the former case, $N$ is fully decoupled from the dark sector heat bath for all temperatures $T$, while in the latter case $N$ is in chemical equilibrium with the dark sector during dark matter freeze-out.}
	\label{fig:reactionrates}
\end{figure}

The cosmological evolution of the dark sector mainly depends on the coupling $g_N$. If it is small enough, the reaction rates for the annihilation processes $\psi \bar \psi \leftrightarrow N \bar N$ and $\phi \phi \leftrightarrow N \bar N$ are smaller than the Hubble rate for all relevant temperatures, implying that $N$ never thermalises with the other states in the dark sector. On the other hand, if $g_N$ is sufficiently large, the process $\phi \phi \leftrightarrow N \bar N$ can lead to equilibration of $\phi$ and $N$ even after the freeze-out of dark matter, until $T \lesssim m_\phi$. Both scenarios are illustrated in Fig.~\ref{fig:reactionrates} by comparing the relevant interaction rates $\Gamma$ to the Hubble rate $H(T)$. In the following section, we will separately discuss the calculation of $\rho_\phi(T)$ and $\rho_N(T)$ in these two different regimes of the coupling strength $g_N$.

\subsection{Cosmological evolution of the energy densities in the dark sector}

Let us start the discussion with the case of sufficiently small couplings $g_N$, corresponding to the case illustrated in the left panel of Fig.~\ref{fig:reactionrates}. More precisely, we define this \emph{small coupling regime} by the requirement that $\Gamma_{\phi\phi \leftrightarrow N \bar{N}}(T) < H(T)$ and $\Gamma_{\psi\bar{\psi} \leftrightarrow N \bar{N}}(T) < H(T)$ for \emph{all} temperatures $T$. In that case, the thermal history of the dark sector can be summarised as follows: both dark matter and the mediator $\phi$ freeze out via $\psi \bar \psi \leftrightarrow \phi \phi$ at a temperature $T_\text{cd}$ corresponding roughly to $T_D(T_\text{cd})\simeq m_\psi/25$~\footnote{As outlined in the previous section, for our numerical calculation we use more precise estimates for the freeze-out temperature following~\cite{Feng:2008mu}.}. At temperatures $T < T_\text{cd}$, the decoupled mediator $\phi$ is only subject to the standard redshift evolution, until it decays efficiently into $N \bar N$ at times $t \gtrsim \tau_\phi$. This situation corresponds precisely to the general setup discussed in detail in section~\ref{sec:evolution}, which we then employ for obtaining the evolution of the energy densities $\rho_\phi(T)$ and $\rho_N(T)$. 

In this discussion, we have implicitly assumed that the initial abundance of sterile neutrinos $N$ prior to the decay of $\phi$ is negligible. However, there is one guaranteed additional source of production of $N$, given by the freeze-in of sterile neutrinos via $\psi \bar{\psi} \rightarrow N \bar{N}$. While this contribution is only relevant for a rather narrow range of $g_N$ within the \emph{small coupling regime}, we nevertheless take it into account in our numerical computations. Details of the freeze-in calculation can be found in appendix~\ref{sec:freezein}.
	
\begin{figure}
	\centering
	\includegraphics[scale=1.1]{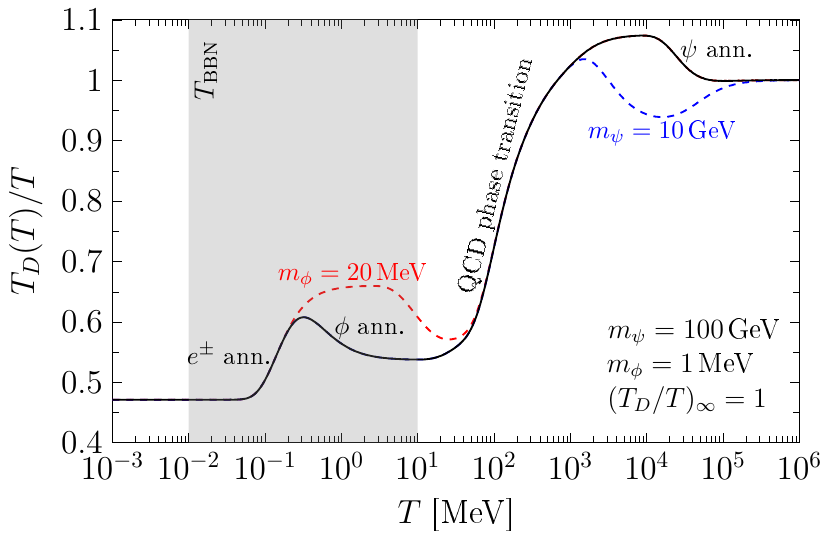}
	\caption{\small Temperature dependence of $T_D(T)/T$ in the \emph{large coupling regime} for exemplary choices of $m_\psi$ and $m_\phi$, fixing $(T_D/T)_\infty = 1$. As indicated in the plot, the disappearance of particles from the SM heat bath leads to a decrease of $T_D(T)/T$, which is most notable during the QCD phase transition. Similarly, the annihilations of $\psi$ and $\phi$ heat up the dark sector relative to the photons, leading to a slight increase of $T_D(T)/T$ at the corresponding temperatures.}
	\label{fig:TempRatio}
\end{figure}
	
Let us now turn to the case of sufficiently large couplings $g_N$, illustrated in the right panel of Fig.~\ref{fig:reactionrates}. This \emph{large coupling regime} is defined by the requirement that $\phi$ and $N$ are in thermal equilibrium during the freeze-out of dark matter; more precisely, we demand $\Gamma_{\phi\phi \leftrightarrow N \bar{N}}(T_\text{ref}) > H(T_\text{ref})$ for $T_\text{ref} \simeq 3m_\psi/(T_D/T)_\infty$. As $\Gamma_{\phi\phi \leftrightarrow N \bar{N}}(T)/H(T)$ is increasing towards smaller values of $T$, this condition ensures chemical equilibrium in the dark sector throughout the dark matter freeze-out process. As in the case of small couplings $g_N$, the dark matter particle freezes out at $T_D(T_\text{cd})\simeq m_\psi/25$, however now the mediator $\phi$ and the sterile neutrino $N$ still follow an equilibrium distribution at $T < T_\text{cd}$ with a temperature $T_D(T)$ determined by entropy conservation in the dark sector. Finally, at temperatures $T$ corresponding to $T_D(T) \lesssim m_\phi$, the equilibrium abundance of $\phi$ gets exponentially suppressed and also the annihilation $\phi\phi \leftrightarrow N \bar{N}$ falls out of equilibrium. We define the corresponding freeze-out temperature $T_{\text{cd}}^{\phi}$ via $\Gamma_{\phi \phi \leftrightarrow N \bar N} (T_D(T_{\text{cd}}^{\phi})) = H(T_{\text{cd}}^{\phi})$, where the Hubble rate $H(T)$ includes both the SM and dark sector energy densities, again properly taking into account the different temperatures of the two sectors. In Fig.~\ref{fig:TempRatio}, we show the evolution of $T_D(T)/T$ for some exemplary choices of $m_\psi$ and $m_\phi$. Importantly, for $(T_D/T)_\infty = 1$ the temperature ratio during BBN is typically $\simeq 0.5-0.7$, leading to a suppression of the additional energy density by roughly one order of magnitude compared to a scenario in which the temperature ratio is $\simeq 1$ for $T \simeq T_\text{BBN}$.

For the calculation of BBN constraints in the \emph{large coupling regime} we need to determine the energy densities of $\phi$ and $N$ at $T \simeq 0.01 - 10\,$MeV, which, depending on the mediator mass and the dark sector temperature ratio, can be either below or above $T_{\text{cd}}^{\phi}$. For $T > T_{\text{cd}}^{\phi}$, the energy densities are simply given by the equilibrium values at the given dark sector temperature $T_D(T)$. When the conversion of $\phi$ and $N$ becomes ineffective at $T_{\text{cd}}^{\phi}$, $\phi$ can decay efficiently into a pair of sterile neutrinos as the inverse reactions are no longer active. For couplings $g_N$ large enough to actually allow for initial equilibration of $N$, the lifetime of $\phi$ is significantly shorter than $t(T_{\text{cd}}^{\phi})$; it is therefore an excellent approximation to assume that all mediators present at $T = T_{\text{cd}}^{\phi}$ instantaneously decay into sterile neutrinos. Taking into account the redshift of the freely propagating sterile neutrinos for $T < T_{\text{cd}}^{\phi}$ we thus finally arrive at
\begin{align}
\rho_D(T) = \rho_\phi(T) + \rho_N(T) & \simeq \begin{cases} \bar{\rho}_\phi (T_D(T)) + \bar{\rho}_N (T_D(T)) &\mbox{for } T > T_{\text{cd}}^{\phi} \\[0.15cm]
\left( \; \bar{\rho}_\phi(T_D(T_{\text{cd}}^{\phi})) + \bar{\rho}_N(T_D(T_{\text{cd}}^{\phi})) \; \right) \\ \quad \quad \times \left( \frac{g_{*s}(T)}{g_{*s}(T_{\text{cd}}^{\phi})} \right)^{4/3} \left( \frac{T}{T_{\text{cd}}^{\phi}} \right)^4 & \mbox{for } T < T_{\text{cd}}^{\phi} \end{cases} \,.
\label{eq:rhoDT_model}
\end{align}	

The BBN bounds are driven by the amount of extra energy density in the dark sector, which will depend on the coupling $g_N$.	
For large values of $g_N$, sterile neutrinos are present in significant amounts in addition to the energy density of $\phi$, whereas for very small $g_N$, $\phi$ becomes non-relativistic before its decay which also increases the dark energy density relative to the visible sector. Consequently, we expect strong bounds from BBN for both very large and very small $g_N$, while intermediate coupling strengths will be somewhat less constrained. We will confirm this expectation quantitatively in the following section.

\subsection{Resulting bounds from BBN}

\begin{figure}
\centering
\hspace*{-1.1cm}
\includegraphics[scale=0.43]{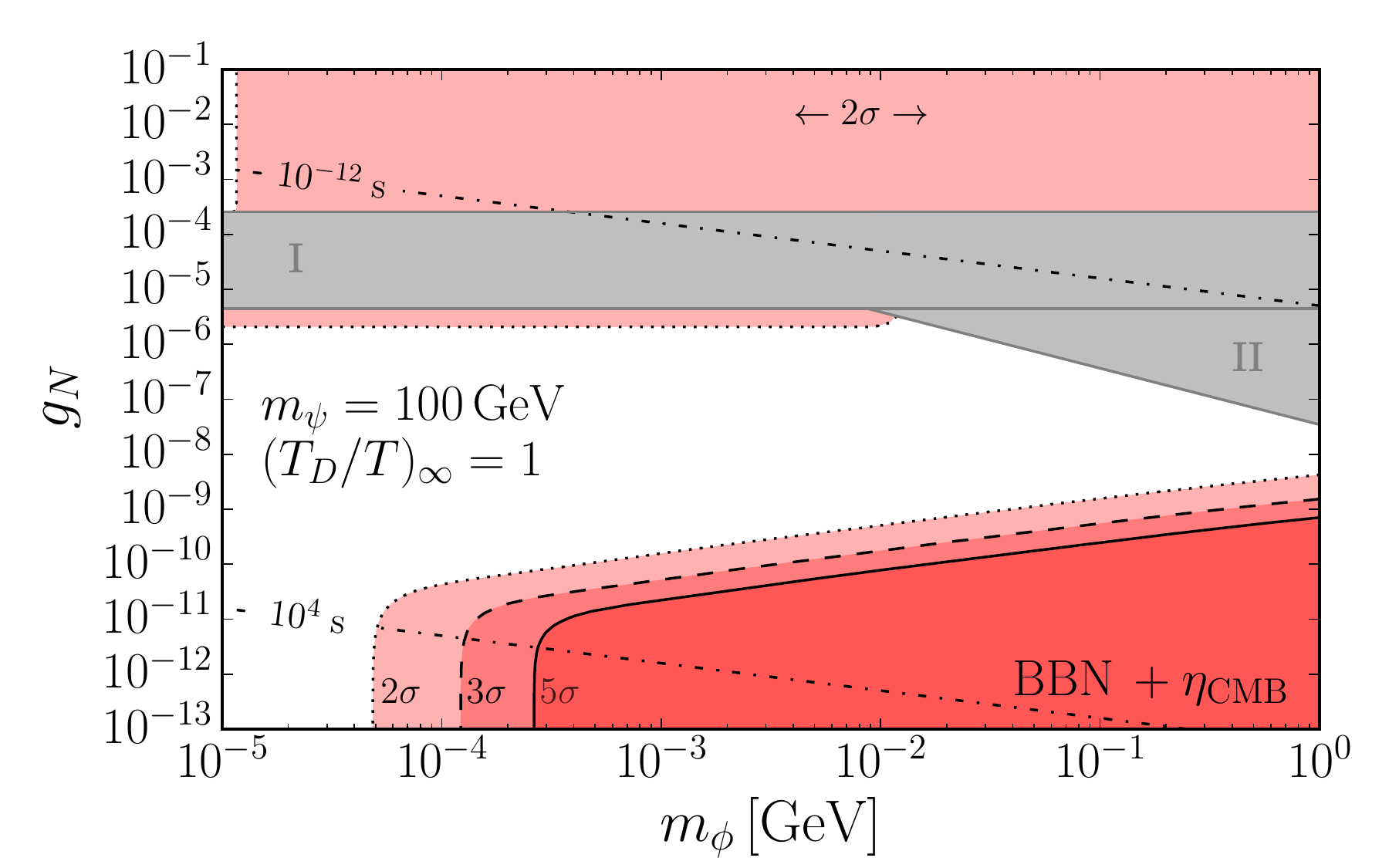}
\includegraphics[scale=0.43]{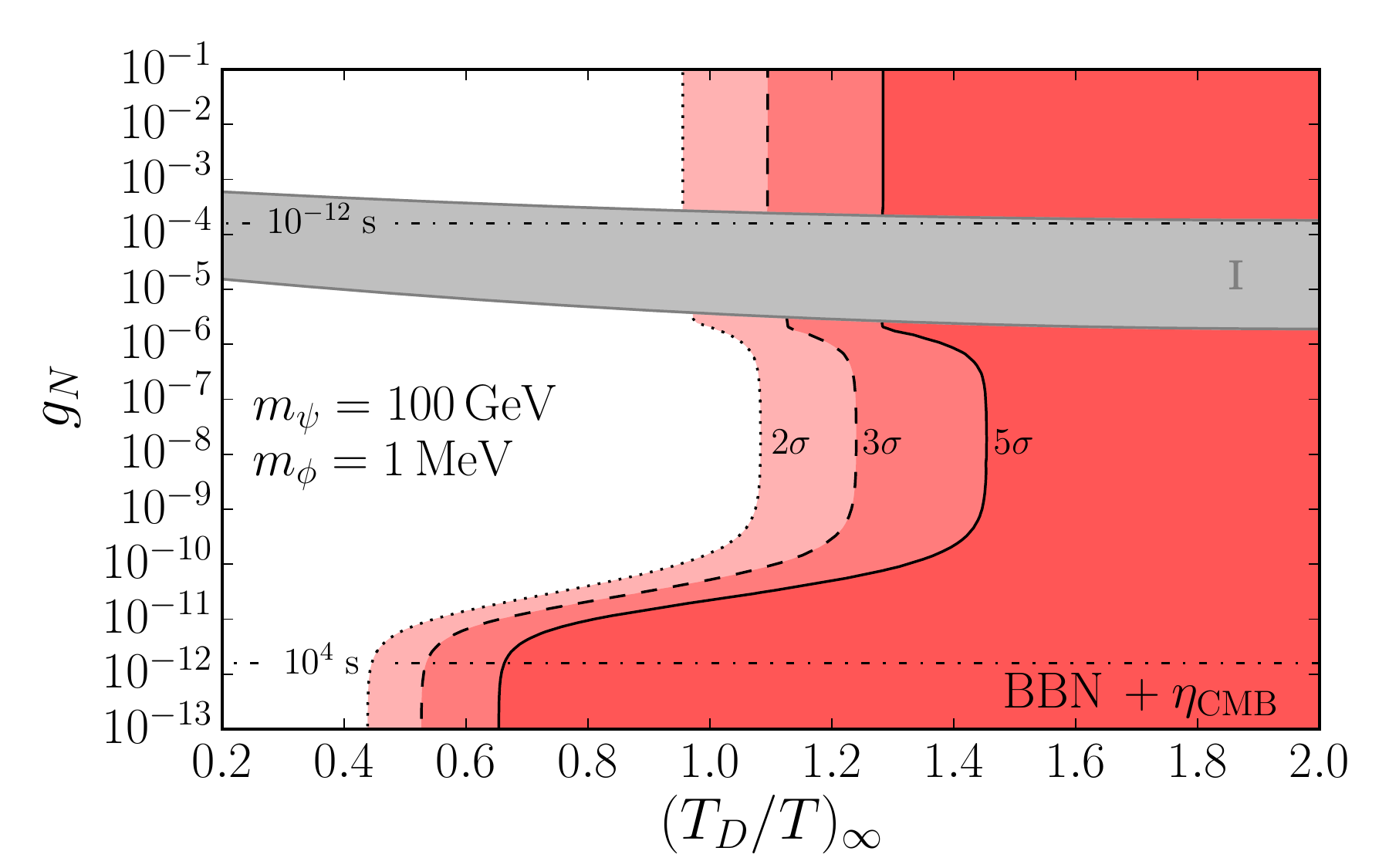}\\
\includegraphics[scale=0.43]{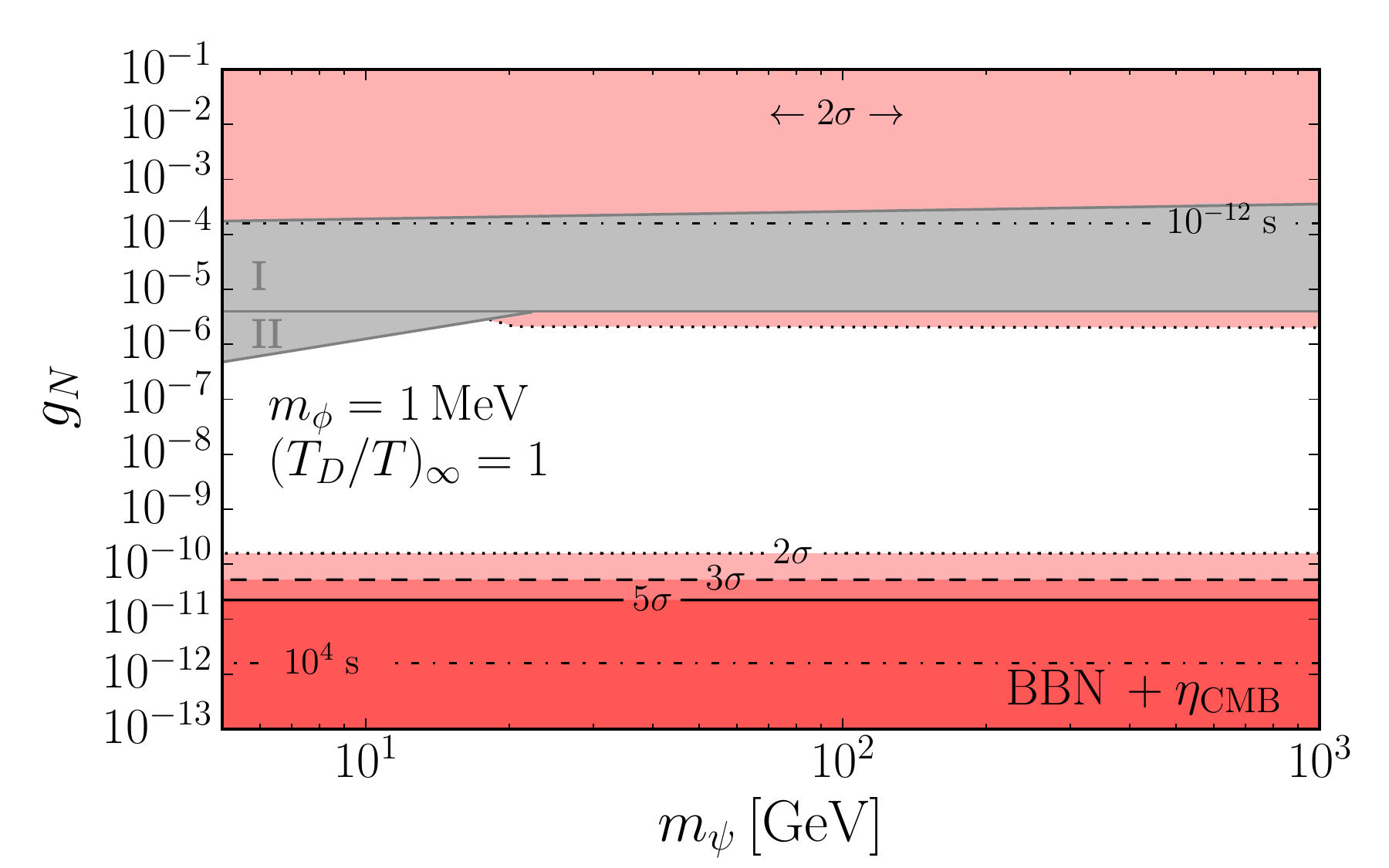}
\caption{\small BBN bounds on the model of dark matter interacting with a light vector mediator as defined in eq.~(\ref{eq:lagrangian}), for different slices in the parameter space. As in Fig.~\ref{fig:general_results}, the red regions enclosed by the black solid, dashed and dotted curves correspond to the parts of parameter space excluded by BBN observations at $5 \sigma$, $3 \sigma$ and $2\sigma$, fixing $\eta_{\rm BBN} = \eta_{\rm CMB}$. The grey shaded regions labelled with I and II correspond to values of $g_N$  fulfilling neither the definition of the large nor of the small coupling regime and hence require a dedicated analysis of the multiparticle Boltzmann equation (see text for details). The black dash-dotted curves indicate contours of fixed lifetimes $\tau_\phi = 10^{-12}\,$s and $10^4\,$s of the mediator $\phi$.}
\label{fig:BBNbounds_model}
\end{figure}

Following the discussion in section~\ref{sec:calc_nuclabundances}, we now compare the calculated nuclear abundances for a given additional energy density $\rho_D(T)$ to the observed values to determine which parts of the parameter space of the model discussed in this section are excluded by BBN at a given confidence level. The results are shown in Fig.~\ref{fig:BBNbounds_model} in various slices of the parameter space. In each case the vertical axis corresponds to the coupling $g_N$ of the vector mediator to sterile neutrinos, which for a given value of $m_\phi$ is in one-to-one correspondence to the lifetime $\tau_\phi$ used in the analogous panels in Fig.~\ref{fig:general_results} (\textit{cf.}~eq.~(\ref{eq:Gammaphi})).

In each of the panels of Fig.~\ref{fig:BBNbounds_model}, the grey shaded region (I) separates the large and small coupling regimes as defined in the previous section: for sufficiently large values of $g_N$ (typically $g_N \gtrsim 10^{-4}$), the sterile neutrino is in chemical equilibrium during DM freeze out, while for much smaller values it never thermalises within the dark sector. The intermediate values of $g_N$ corresponding to region (I) indicate the range of parameters which neither corresponds to the small nor to the large coupling regime: in this case, the interaction between $\phi$ and $N$ is not efficient enough to bring the sterile neutrinos into equilibrium for all temperatures relevant to the dark matter freeze-out process, but it is nevertheless sufficiently large such that $\psi \bar \psi \leftrightarrow N \bar N$ leads to equilibration of $N$ at least for some temperatures. This part of the parameter space thus corresponds to the transition between the freeze-in and freeze-out of $N$, and a detailed discussion is beyond the scope of this work. Furthermore, in Fig.~\ref{fig:BBNbounds_model} we do not consider the region of parameter space corresponding to $\left< \Gamma_{\phi \rightarrow N \bar{N}} \right>(T_\text{cd})/H(T_\text{cd}) \lesssim 0.1$, labeled by (II). In this case, the mediator $\phi$ decays already before the decoupling of the dark matter particle; as for those ranges of parameters $\phi$ is not in thermal contact with any other particle of the dark sector, the abundance of the (decaying) particle $\phi$ does not follow an equilibrium distribution and hence one cannot apply the standard freeze-out calculation. We leave a closer inspection of those regions of the parameter space for future work.

The BBN bounds in the {\it small coupling regime} in each panel of Fig.~\ref{fig:BBNbounds_model} directly map onto the general results presented in section~\ref{sec:evolution}. More precisely, a given dark matter mass $m_\psi$ and an initial temperature ratio $(T_D/T)_\infty$ correspond to exactly one pair of decoupling temperatures $T_\text{cd}$ and $T_D(T_\text{cd})$, and vice versa. Hence, the dependence of the bounds in this part of the parameter space on the various model parameters can be directly understood from the discussion in section~\ref{sec:generalBBN_bounds}. In particular, BBN poses very stringent constraints once the temperature ratio of the dark and visible sector is sufficiently large, or if the mediator becomes non-relativistic prior to its decay at small $g_N$ and large $m_\phi$. 

For values of $g_N$ above the grey bands in Fig.~\ref{fig:BBNbounds_model}, i.e.~in the {\it large coupling regime}, the additional energy density associated to the thermalised sterile neutrinos $N$ leads to somewhat stronger constraints from BBN compared to intermediate coupling strenghts. In particular, for an initial temperature ratio $(T_D/T)_\infty=1$, all of the large coupling regime is in tension with BBN observations at $\sim 2\sigma$ or more.

Lastly, let us discuss the impact of the freeze-in process $\psi \bar \psi \rightarrow N \bar N$ on the bounds shown in Fig.~\ref{fig:BBNbounds_model}. This production mechanism is only relevant for the largest values of $g_N$ within the small coupling regime: in this case, the rate of production of $N$ is still too small in order to allow for equilibration, but already large enough such that it leads to a population of sterile neutrinos which is comparable to the thermal abundance of $\phi$. This is clearly visible in all panels of Fig.~\ref{fig:BBNbounds_model}: the bounds from BBN get stronger for $g_N$ slightly below the intermediate region (I). In fact, the right upper panel suggests a smooth transition between the small and large coupling regimes once the freeze-in contribution is taken into account. It is hence reasonable to assume that the BBN bounds in this part of parameter space approximately interpolate between the two different regimes; nevertheless, we conservatively mask out these ranges of parameters in our analysis.

\subsection{Confronting self-interactions and late kinetic decoupling with BBN bounds}

\begin{figure}
\begin{center}
\hspace*{-1.0cm}
\includegraphics[scale=0.93]{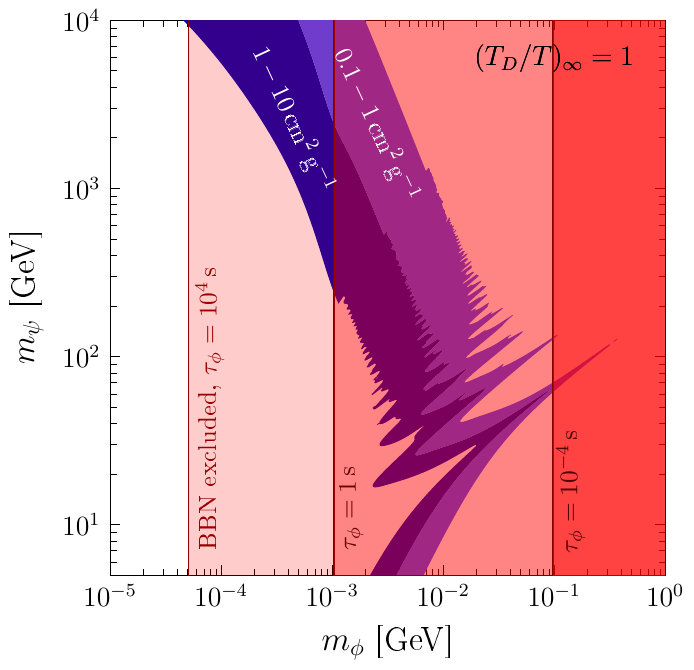}
\hspace*{0.3cm}
\includegraphics[scale=0.93]{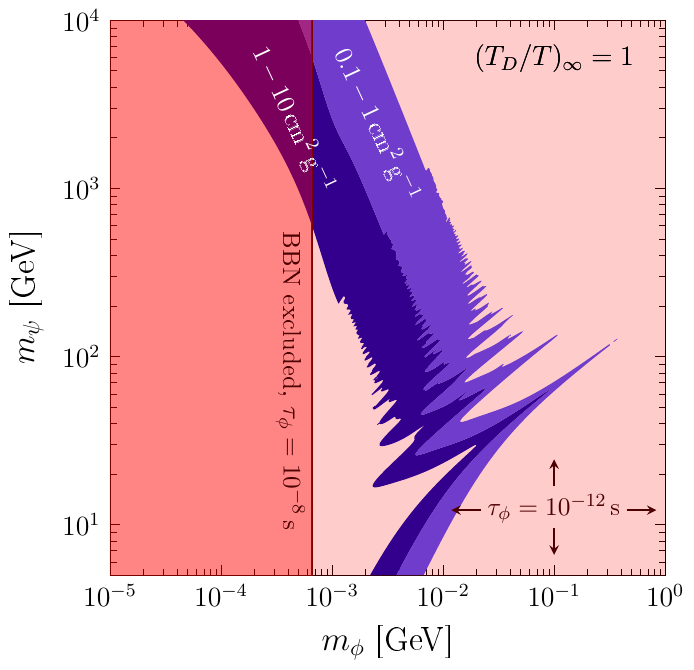}
\end{center}
\caption{\small $2\sigma$ exclusion limits from BBN in the parameter space spanned by $m_\phi$ and $m_\psi$, for various lifetimes $\tau_\phi$ and an initial temperature ratio $(T_D/T)_\infty = 1$. In addition we show the regions in parameter space for which the DM self scattering cross section $\sigma_T/m_\psi$ is in the range $0.1-10$  cm$^2$/g, as required by solving the small-scale problems of standard cold dark matter. For better visibility we show large lifetimes in the \textit{left panel} and small lifetimes in the \textit{right panel}.}
\label{fig:SIDM_results}
\end{figure}

As already mentioned previously, the model of dark matter coupled via an MeV-scale vector mediator to sterile neutrinos naturally leads to strong self-interactions of dark matter. We show in Fig.~\ref{fig:SIDM_results} the regions of parameter space spanned by $m_\phi$ and $m_\psi$ leading to a momentum transfer cross section $\sigma_T/m_\psi$ of $0.1-1\,\text{cm}^2/\text{g}$ (light blue band) and $1-10 \, \text{cm}^2/\text{g}$ (dark blue band) at dwarf scales, i.e.~$v \simeq 30\,\text{km}/\text{s}$. Here we have set the initial temperature ratio $(T_D/T)_\infty=1$, and again fixed $g_\psi$ by the requirement of correctly reproducing the observed amount of dark matter via thermal freeze-out\footnote{In the calculation of $\sigma_T/m_\psi$ we follow~\cite{Kahlhoefer:2017umn}; in particular, we fully take into account quantum indistinguishability in the scattering process and consequently adopt the definition of $\sigma_T$ involving an integration over $\text{d}\sigma/\text{d}\Omega$ weighted by $1-|\text{cos} \, \theta|$.}. We then superimpose the $2\sigma$ bounds from BBN for fixed values of the lifetime $\tau_\phi$, concretely for $\tau_\phi = 10^{-4}, 1$ and $10^4\,$s in the left panel and for $\tau_\phi = 10^{-8}$ and $10^{-12}\,$s in the right panel. Clearly, depending on the lifetime of the mediator (or equivalently the coupling $g_N$) BBN may or may not exclude values of the dark matter and mediator mass leading to the desired self interaction cross section. Furthermore, increasing (decreasing) the initial temperature ratio of the dark and visible sector would significantly strengthen (weaken) the bounds from BBN on this model of self-interacting dark matter, as can be seen e.g.~from the right upper panel in Fig.~\ref{fig:BBNbounds_model}.

\begin{figure}
\begin{center}
\includegraphics[scale=1.3]{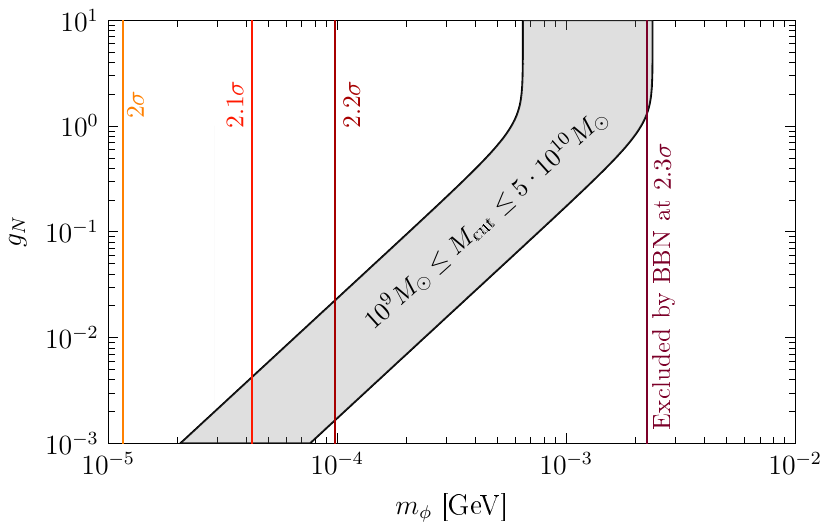}
\end{center}
\caption{\small BBN bounds in the part of parameter space potentially solving the missing satellite problem via late kinetic decoupling of dark matter, choosing for definiteness $m_\psi = 2\,$TeV and $(T_D/T)_\infty = 1$. The grey shaded region enclosed by the black solid curves corresponds to kinetic decoupling temperatures $T_\text{kd}$ leading to a small-scale cutoff $M_\text{cut}$ in the power spectrum of matter density perturbations which is in the right range in order to solve the missing satellite problem.}
\label{fig:LKD_results}
\end{figure}

Finally, in Fig.~\ref{fig:LKD_results} we confront the regions of parameter space leading to the desired cutoff in the power spectrum of matter density perturbations with the bounds from BBN. In the model discussed in this work, dark matter remains in kinetic equilibrium via the scattering process $\psi N \leftrightarrow \psi N$, as discussed in more detail in e.g.~\cite{Bringmann:2013vra,Binder:2016pnr,Bringmann:2016ilk}. This process falls out of equilibrium at a photon temperature~\cite{Bringmann:2013vra}
 \begin{align}
 T_{\text{kd}}^{\psi} \simeq \frac{62\,\text{eV}}{(g_\psi g_N)^{1/2}} \cdot \left( \frac{T_{\text{kd}}^{\psi}}{T_D(T_{\text{kd}}^{\psi})} \right)^{3/2} \cdot \left( \frac{m_\psi}{\text{TeV}} \right)^{1/4} \cdot \frac{m_\phi}{\text{MeV}} \,,
 \label{eq:Tkd}
 \end{align}
leading to a cutoff $M_\text{cut}$ in the power spectrum of matter density perturbations given by~\cite{Bringmann:2013vra}
 \begin{align}
 M_\text{cut} \simeq 1.7 \cdot 10^8 M_\odot \cdot \left(\frac{T_{\text{kd}}^{\psi}}{\text{keV}} \right)^{-3} \,.
\label{eq:Mcut}
 \end{align}
The grey region in Fig.~\ref{fig:LKD_results} then corresponds to the model parameters leading to interesting cutoff masses~\cite{Bringmann:2013vra} $10^9 M_\odot \lesssim M_\text{cut} \lesssim 5 \cdot 10^{10} M_\odot$. As apparent from the plot,  there is a $\sim2\sigma$ tension between this solution of the missing satellite problem and the bounds on the resulting extra energy density from BBN~\footnote{Notice that even in the large coupling regime $\Delta N_\text{eff}(T)$ has a non-trivial time dependence during BBN (provided that $m_\phi \sim T_\text{BBN}$), even though it is typically less pronounced than for the example shown in the right panel of Fig.~\ref{fig:rhoT_Nefft_modelindependent}. We fully take this into account in our derivation of the BBN bounds. Let us also mention that for mediator masses $m_\phi \lesssim 10^{-5}\,$MeV the bound from BBN stays constant at the level of $\simeq 2 \sigma$; however, this region of parameter space is anyway excluded for all relevant dark matter masses by giving rise to too large self-interaction cross sections (\emph{cf.}~Fig.~\ref{fig:SIDM_results}), and is hence omitted from the plot.}. 

In Fig.~\ref{fig:LKD_results}, we have fixed the dark matter mass to $m_\psi = 2\,$TeV and the initial temperature ratio $(T_D/T)_\infty=1$, so let us briefly discuss the impact of those choices on our conclusions. First, as shown in the lower panel of Fig.~\ref{fig:BBNbounds_model}, the bounds from BBN are largely independent of the dark matter mass as long as $m_\psi \gtrsim 20 \cdot \Lambda_\text{QCD} \simeq 5\,$GeV. As shown in~\cite{Bringmann:2013vra} (see in particular Fig.~2 therein), also the region in parameter space corresponding to the interesting range for $M_\text{cut}$ is rather insensitive to $m_\psi$. Consequently, the level of tension of BBN with this solution of the missing satellite problem does not strongly vary with the dark matter mass. On the other hand, we have discussed several times that the bounds from BBN are significantly weakened for a smaller initial temperature ratio $(T_D/T)_\infty$. In that case also the region in parameter space leading to $10^9 M_\odot \lesssim M_\text{cut} \lesssim 5 \cdot 10^{10} M_\odot$ would shift to smaller values of $m_\phi$ (cf.~eqs.~(\ref{eq:Tkd}) and~(\ref{eq:Mcut})), but as the bound from BBN is not very sensitive to the mediator mass in the large coupling regime, the overall tension of BBN with the explanation of the missing satellite problem within this model is significantly reduced for $(T_D/T)_\infty$  sufficiently smaller than one.

\section{Conclusions}
\label{sec:conclusions}
 
The remarkable agreement between the observed primordial abundances of light nuclei and the predictions based on the Standard Model particle content puts severe constraints on any additional energy density present at $T_\text{BBN} \simeq (0.01 - 10)\,$MeV. This applies in particular to a fully decoupled dark sector, i.e.~a collection of particles which have no (or highly suppressed) non-gravitational interactions with particles of the Standard Model. Such scenarios are notoriously difficult to constrain using laboratory experiments, so that cosmological probes such as BBN remain the only hope to either constrain or give evidence for models of this type.
 
This idea has been explored extensively in the two limits of a decaying particle which is either non- or ultra-relativistic during BBN. Motivated by dark matter scenarios involving a MeV-scale mediator $\phi$ (i.e.~$m_\phi \simeq T_\text{BBN}$), in the first part of this work we close this gap by performing a largely model-independent study of BBN bounds on a particle with a mass in the MeV range decaying into other light states of the dark sector. To this end, we first study in detail the cosmological evolution of the energy densities in the dark sector fully taking into account the finite mass of $\phi$. We find that the corresponding $\Delta N_\text{eff}(T)$ can in general have a strong temperature dependence during BBN, thus not allowing for a simple comparison with a bound based on a constant number of additional neutrinos. Instead, we employ the public code \texttt{AlterBBN} in order to derive the abundances of light elements point-by-point in the parameter space spanned by the particle mass, lifetime and chemical decoupling temperature, as well as the ratio of the temperatures in the dark and visible sector. We then compare those to the latest measurements of primordial abundances. Taking into account both the observational errors as well as systematic uncertainties on the nuclear cross sections we explore which part of the parameter space is excluded by BBN at a given confidence level. 

Our results imply strong bounds from BBN on such dark sector scenarios, provided that (1) the temperature ratio of the dark and visible sector is of order unity or larger, and/or (2) the particle $\phi$ is non-relativistic for a sufficiently long time before it decays, and thus has an increased energy density compared to the (radiation dominated) Standard Model contribution. We refer to Fig.~\ref{fig:general_results} for a more detailed summary of our results in various slices of the parameter space. As a limiting case of our approach, we furthermore update and improve the bounds on fully non-relativistic particles decaying into dark sector states as presented in~\cite{1988ApJ...331...33S}.

In the second part of this work, we apply our general discussion of BBN bounds to a specific model involving a dark matter candidate $\psi$ interacting with light fermionic states $N$ via a vector mediator $\phi_\mu$ with a mass in the MeV range. Such a setup generically leads to a significant self-interaction cross section of dark matter as well as late kinetic decoupling, leading to a cutoff in the power spectrum of matter density fluctuations. It has therefore been proposed as a solution to {\it all} small scale structure problems of the collisionless cold dark matter paradigm on small scales, including the missing satellite problem. Carefully taking into account the cosmological evolution of the dark sector in different regimes of the coupling strength we derive  bounds on the model from BBN observations. We find viable regions in parameter space which lead to the relevant dark matter self-interactions, while regions which address the missing satellite problem with a cutoff in the matter power spectrum are in $\sim 2 \sigma$ tension with BBN.

\acknowledgments

We thank Tobias Binder, Torsten Bringmann, Camilo Garcia Cely and Felix Kahlhoefer for helpful discussions. This work is supported by the German Science Foundation
(DFG) under the Collaborative Research Center~(SFB) 676 Particles, Strings and the Early Universe as well as the ERC Starting Grant `NewAve' (638528).

\appendix
\section{Impact of kinetic decoupling}
\label{sec:impact_kineticdecoupling}

In section~\ref{sec:evolution} we have assumed that after chemical decoupling $\phi$ is also kinetically decoupled from the dark sector heat bath, i.e.~$T_\text{cd} = T_\text{kd}$. The actual value of $T_\text{kd}$ is model-dependent; however, we show in this appendix that it is largely irrelevant for the temperature evolution of the energy density $\rho_\phi$ and hence for the constraints imposed on the scenario by BBN.

After chemical decoupling the Boltzmann equation for $\phi$ in general is given by
\begin{align}
H(T)\sqrt{m_\phi^2 + p^2}\left( \frac{T}{1+\Delta_{*s}(T)} \frac{\partial}{\partial T} + p \frac{\partial}{\partial p} \right) f_\phi(T,p) = \frac{m_\phi}{\tau_\phi} f_\phi(T, p) -  \mathcal{C}_{\text{elastic}}[f_\phi]
\label{eq:Boltzmann_eqn_phi_C}
\end{align}
with $\mathcal{C}_\text{elastic}[f_\phi]$ being the collision operator associated to any elastic (i.e.~not number-changing) scattering process of $\phi$. As long as the bosonic mediator is in kinetic equilibrium with a heat bath of temperature $T_D(T)$, its phase space density is given by
\begin{align}
 f_\phi (E,T) \big|_{T > T^{\text{(kd)}}} = \left[\exp \left( \frac{E - \mu_\phi(T)}{ T_D(T) }\right) - 1 \right]^{-1} \,.
\label{eq:fphi_kinetic_eq}
\end{align}
Consequently, calculating $f_\phi(E, T)$ is equivalent to finding the two unknown functions $\mu_\phi(T)$ and $T_D(T)$. For simplicity, let us consider the case of a stable mediator, i.e.~in the following we take the limit $\tau_\phi \rightarrow \infty$. Integrating eq.~(\ref{eq:fphi_kinetic_eq}) over all momenta and using $\int \text{d}^3 p \, E^{-1}\mathcal{C}_\text{elastic}[f_\phi] = 0$~\cite{Kolb:1990vq}, we recover particle conservation in the form
\begin{equation}
\frac{\text{d}n_\phi(T)}{\text{d}T} - 3\frac{1+\Delta_{*s}(T)}{T}n_\phi(T) = 0\;\,.
\label{eq:particlenumbercons_ds}
\end{equation}
Moreover, entropy conservation (in case of a stable mediator) in the dark sector implies
\begin{equation}
0 = T_D\text{d}S = \text{d}(R^3\varrho_\phi) + p_\phi \text{d}R^3 - \mu_\phi (R^3n_\phi)\;\,,
\end{equation}
leading to
\begin{equation}
\frac{\text{d}\varrho_\phi(T)}{\text{d}T} - 3\frac{1+\Delta_{*s}(T)}{T}\left(\varrho_\phi(T) + p_\phi(T)\right) = 0\;\,.
\label{eq:entropycons_ds}
\end{equation}
Eqs.~(\ref{eq:particlenumbercons_ds}) and~(\ref{eq:entropycons_ds}) form a set of integro-differential equations for the two functions $\mu(T)$ and $T_D(T)$ which we solve numerically using the initial conditions $\mu(T_\text{cd}) = 0$ and $T_D(T_\text{cd}) = T_{D, \text{cd}}$.

\begin{figure}
\begin{center}
\includegraphics[scale=0.86]{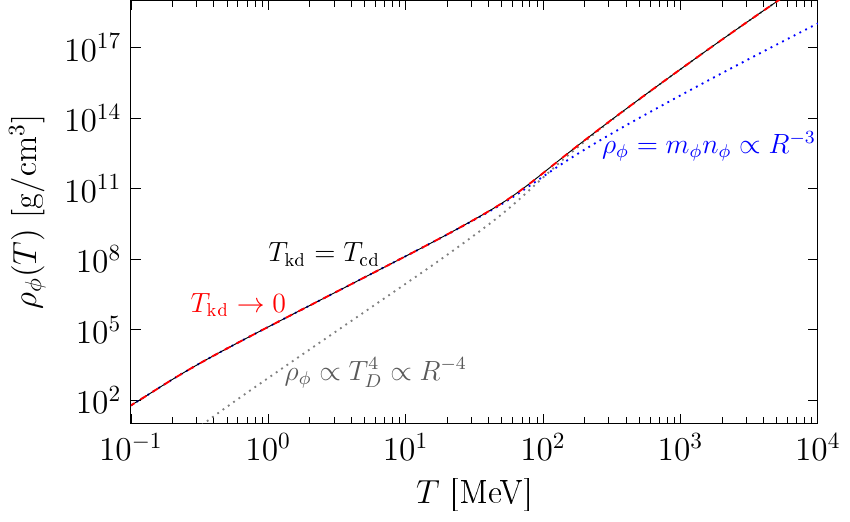}
\includegraphics[scale=0.86]{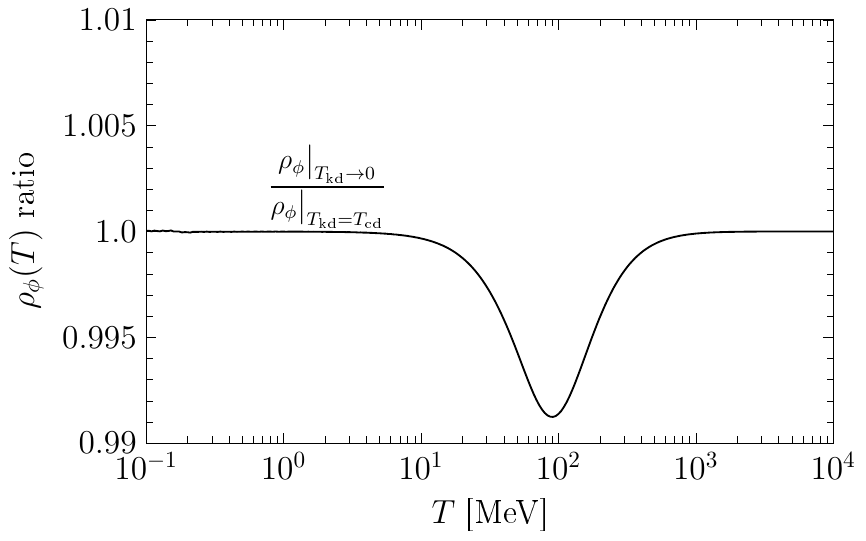}
\end{center}
\caption{\small \textit{Left panel:} evolution of $\rho_\phi(T)$ for two different choices of the kinetic decoupling temperature $T_\text{kd}$ (see text for details). Here we assume $m_\phi = 100\,$MeV, $T_{\text{cd}} = 10\,$GeV, $T_D (T_{\text{cd}}) = 5\,$GeV as well as a stable mediator. \textit{Right panel:} relative difference of $\rho_\phi(T)$ when assuming $T_\text{kd} \rightarrow 0$ and $T_\text{kd} \rightarrow \infty$.}
\label{fig:ImpactOfKineticDecoupling}
\end{figure}

We illustrate the impact of $T_\text{kd}$ on our results in Fig.~\ref{fig:ImpactOfKineticDecoupling}. Here we show the temperature evolution of $\rho_\phi(T)$ for the example case of $m_\phi = 100\,$MeV,  $T_{\text{cd}} = 10\,$GeV, $T_D(T_\text{cd}) = 5\,$GeV.  The black solid curve shows $\rho_\phi(T)$ following from the formalism in section~\ref{sec:evolution} under the assumption that $T_\text{kd} = T_\text{cd}$, i.e.~that all of the evolution after chemical decoupling simply follows from red-shifting the momenta. On the other hand, the red dashed curve corresponds to the other extreme where $T_\text{kd} \rightarrow 0$, i.e.~the mediator remains in kinetic equilibrium for all temperatures of interest. The right panel in the same figure shows the relative difference of the two energy densities. It follows that both for $T \gg m_\phi$ and $T \ll m_\phi$ both assumptions lead to the same $\rho_\phi(T)$, and that for $T \sim m_\phi$ the difference between the two energy densities is below $1 \%$.

This can be understood as follows: if $T_\text{kd} = T_\text{cd}$, all the evolution of the phase space density of $\phi$ follows from redshifting the momenta according to $p \propto 1/R$. For $T \gg m_\phi$, where $\phi$ is essentially massless, this simply amounts to set $T_D \propto 1/R$ and $\mu_\phi = 0$. On the other hand, if $\phi$ is in kinetic equilibrium during $T \gg m_\phi$, the requirement that $n_\phi \propto 1/R^3$ and $s_\phi \propto 1/R^3$ leads to the same behaviour of $T_D$ and $\mu_\phi$, and thus to the same energy density $\rho_\phi(T)$.  Moreover, for $T \ll m_\phi$ the energy density of the mediators is simply given by $\rho_\phi(T) \simeq m_\phi n_\phi(T)$, where $n_\phi \propto 1/R^3$ is fixed by particle-number conservation. Hence, also in this regime the energy density of $\phi$ is independent of the kinetic decoupling temperature. 

Consequently,  $T_\text{kd}$ can only affect the evolution of $\rho_\phi(T)$ in the semi-relativistic regime where $T \sim m_\phi$. Indeed, it is well-known that redshifting a thermal distribution of semi-relativistic particles in general does not lead to a thermal distribution at later temperatures~\cite{Kolb:1990vq}, explaining the difference between the two cases considered in Fig.~\ref{fig:ImpactOfKineticDecoupling}. However, we find that for essentially all choices of parameters the difference in the regime $T \sim m_\phi$ is below the percent level (see right panel of Fig.~\ref{fig:ImpactOfKineticDecoupling}). Hence, we can safely neglect the impact of $T_\text{kd}$ on our BBN constraints, justifying our procedure to set $T_\text{kd} = T_\text{cd}$.

\section{Freeze-in of the sterile neutrinos}
\label{sec:freezein}

In this appendix we provide the details of our derivation of the abundance of sterile neutrinos $N$ resulting from the freeze-in process $\psi \bar \psi \rightarrow N \bar N$. For all parameters of interest in this work, the freeze-in of $N$  concludes before the freeze-out of $\phi$ and $\psi$ at approximately $T_D \sim m_\psi/10$, i.e.~long before the decay of the mediator. Hence, the mutual influence of freeze-in and decay is negligible, implying that it is possible to write down an independent Boltzmann equation for the freeze-in process given by
\begin{equation}
\frac{\text{d}\rho_N(T)}{\text{d}T} - 4\frac{1+\Delta_{*s}(T)}{T} \rho_N(T) = - \frac{1+\Delta_{*s}(T)}{H(T)T} \left[ \left< E_{\psi\bar{\psi}} \sigma v_{\mathrm{M\o{}l}} \right> \bar{n}_{\psi}^2  \right](T_D(T))\;, 
\label{rhoN_freezein}
\end{equation}

\noindent with $\rho_N(T \rightarrow \infty) = 0$ and the thermal average~\cite{Gondolo:1990dk}
\begin{align}
\left[ \left< E_{\psi\bar{\psi}} \sigma v_{\mathrm{M\o{}l}} \right> \bar{n}_{\psi}^2  \right]&(T_D(T)) = \nonumber \\
&\int \frac{2 \,\text{d}^3p_1}{(2 \pi)^3} \frac{2 \,\text{d}^3p_2}{(2 \pi)^3} (E_1 + E_2) \frac{\sigma_{\psi\bar{\psi}\rightarrow N\bar{N}}F}{E_1E_2}\; \bar{f}_{\psi}(T_D(T), p_1) \bar{f}_{\bar{\psi}}(T_D(T),p_2)\;\,,
\label{Esv_integral}
\end{align}
where we defined $E_{\psi \bar{\psi}} \equiv E_1 + E_2$ and $\sigma v_{\mathrm{M\o{}l}} = \sigma_{\psi\bar{\psi}\rightarrow N\bar{N}}F/E_1E_2$. The solution to eq.~(\ref{rhoN_freezein}) is given by
\begin{align}
\rho_N(T) = \left[g_{*s}(T)^{\sfrac13}T\right]^4  \int_{T}^{\infty}  \frac{1}{\left[g_{*s}(\lambda)^{\sfrac13}\lambda\right]^4} \cdot \left[ \left< E_{\psi\bar{\psi}} \sigma v_{\mathrm{M\o{}l}} \right> \bar{n}_{\psi}^2  \right](T_D(\lambda)) \cdot \frac{1+\Delta_{*s}(\lambda)}{H(\lambda)\lambda}\; \text{d}\lambda\;\,.
\label{freeze_in_solution}
\end{align}
As usual, $T_D(T)$ follows from entropy conservation in the dark sector, while the product of the cross section and the flux factor $F$ is given by
\begin{equation}
\sigma_{\psi\bar{\psi}\rightarrow N\bar{N}}F = \frac{g_\psi^2 g_N^2}{12 \pi} \frac{s+2m_\psi^2}{2s}\;\,.
\end{equation}

\bibliography{refs}
\bibliographystyle{ArXiv}

\end{document}